\documentclass[journal]{IEEEtran}

\normalsize
\ifCLASSINFOpdf
\else
\fi

\usepackage{amsthm}
\usepackage{multicol}
\usepackage{graphicx}
\usepackage{amssymb, amsmath}
\usepackage{xcolor}
\usepackage{cite}
\usepackage{hyperref}
\usepackage{enumerate}
\usepackage{esint}
\usepackage{balance}
\usepackage{textcomp}

\DeclareMathOperator\erf{erf}

\theoremstyle{plain}

\providecommand{\definitionname}{Definition}

\makeatother

\theoremstyle{plain}
\newtheorem{thm}{\protect\theoremname}
\providecommand{\theoremname}{Theorem}

\makeatother

\usepackage{nomencl}
\makenomenclature

\begin{document}

\title{All-Optical Cochlear Implants}

\author{Stylianos~E.~Trevlakis,~\IEEEmembership{Student~Member,~IEEE,}
        Alexandros--Apostolos~A.~Boulogeorgos,~\IEEEmembership{Senior Member,~IEEE,}
        Nestor~D.~Chatzidiamantis,~\IEEEmembership{Member,~IEEE,}
        and~George~K.~Karagiannidis,~\IEEEmembership{Fellow,~IEEE} 
        \vspace{-0.7cm}
\thanks{The authors are with the Department of Electrical and Computer Engineering, Aristotle University of Thessaloniki, Thessaloniki, 54124 Greece. e-mails: \{trevlakis, nestoras, geokarag\} @auth.gr, al.boulogeorgos@ieee.org.}%
\thanks{Alexandros--Apostolos~A.~Boulogeorgos is also with the Department of Digital Systems, University of Piraeus, Piraeus 18534, Greece.}%
\thanks{The work of S. E. Trevlakis was supported by the Hellenic Foundation for Research and Innovation (H.F.R.I.) in the context of the `2nd Proclamation of Scholarships from ELIDEK for PhD Candidates' (G.A. No. 16788).}
\thanks{Manuscript received -, 2020; revised -, 2020.}
}

\markboth{IEEE Transactions on Molecular, Biological, and Multi-Scale Communications,~Vol.~-, No.~-, -~2020}%
{Shell \MakeLowercase{\textit{et al.}}: Bare Demo of IEEEtran.cls for IEEE Journals}

\maketitle

\begin{abstract}
In the present work, we introduce a novel cochlear implant (CI) architecture, namely all-optical CI (AOCI), which directly converts acoustic to optical signals capable of stimulating the cochlear neurons. First, we describe the building-blocks (BBs) of the AOCI, and explain their functionalities as well as their interconnections. Next, we present a comprehensive system model that incorporates the technical characteristics and constraints of each BB, the transdermal-optical-channel particularities, i.e. optical path-loss and external-implanted device stochastic pointing-errors, and the cochlear neurons biological properties. Additionally, in order to prove the feasibility of the AOCI architecture, we conduct a link-budget analysis that outputs novel closed-form expressions for the instantaneous and average photon flux that is emitted on the cochlear neurons. Likewise, we define three new key-performance-indicators (KPIs), namely probability of hearing, probability of false-hearing, and probability of neural damage. The proposed theoretical framework is verified through respective simulations, which not only quantify the efficiency of the proposed architecture, but also reveal an equilibrium between the optical transmission power and the patient's safety, as well as the AOCI BBs specifications. Finally, it is highlighted that the AOCI approach is greener and safer than the conventional~CIs.
\end{abstract}

\begin{IEEEkeywords}
Average photon flux, biomedical communications, cochlear implants, feasibility study, link-budget, optogenetics, transdermal optical communications. \vspace{-0.4cm}
\end{IEEEkeywords}

\IEEEpeerreviewmaketitle

\section*{Nomenclature}
\addcontentsline{toc}{section}{Nomenclature}
\begin{IEEEdescription}[\IEEEusemathlabelsep\IEEEsetlabelwidth{$V_1,V_2,V_3$}]
\item[ABR] Auditory brainstem response
\item[AOCI] All-optical cochlear implant
\item[CCI] Conventional cochlear implant
\item[CI] Cochlear implant
\item[CL] Coupling lens
\item[DSP] Digital signal processing
\item[FBG] Fiber Bragg grating
\item[GL] Guided lens
\item[KPI] Key performance indicator
\item[LD] Laser diode
\item[LED] Light emitted diode
\item[LS] Light source
\item[MEM] Microelectromechanical
\item[MPE] Maximum permissible exposure
\item[OF] Optical fiber
\item[PDF] Probability density function
\item[PMF] Probability mass function
\item[RF] Radio frequency
\item[SNR] Signa-to-noise-ratio
\item[TOL] Transdermal optical link
\end{IEEEdescription}

\section{Introduction}
Over half a million patients worldwide, suffering from mild-to-severe hearing loss, regain their speech perception with the aid of cochlear implants (CIs)~\cite{dombrowski2018toward}. Conventional CIs (CCIs) consist of an external device that converts the audio into radio frequency (RF) signals and emits them to an implanted device. The latter processes the received signals in order to convert them into electrical ones, which stimulate the acoustic nerve. The main bottlenecks of the CCIs are twofold: (i) The poor coding of spectral information due to the wide spread of electric current from each electrode, which results in broad excitation of the cochlear neurons; and (ii) The bandwidth scarcity on the RF spectrum that significantly constraints the achievable data rate; hence, the faithful reproduction of the neural stimulus~\cite{wilson2008implants,zeng2008implants,wilson2017modern,Moser2020}.

\vspace{-0.4cm}
\subsection{State Of The Art and Motivation}
In order to surpass the electrical stimulation constraints, a great amount of research effort has been directed toward alternative neural stimulation mechanics. One of the most promising concepts is the excitation of the neurons via optical signals~\cite{Hernandez2014,wrobel2018optogenetic,mager2018high,keppeler2018ultrafast,Xu2018,dombrowski2018toward}. This approach is widely known as optogenetic stimulation and utilizes optical radiation in the wavelengths from $450$ to $600\text{ }\mathrm{nm}$. Optogenetic interfaces are characterized by increased spatial resolution in comparison with electrical ones, and present higher tolerance for unwanted stimulation artifacts that interfere with the desirable stimulation~\cite{Hart2019}. Aspired by this, in~\cite{Hernandez2014}, the authors experimentally verified that optogenetic cochlear stimulation achieves increased temporal fidelity with low-light intensities. This reveals that optogenetics can be used to develop CIs with improved restorative capabilities. Likewise, in~\cite{wrobel2018optogenetic,mager2018high,keppeler2018ultrafast}, the narrow-light-spread in the cochlea was revealed, and it was indicated that excitation can be achieved with optogenetic stimulation with emitter intensities in the order of $\mathrm{mW}$. Note that, the CCI requires transmission power of approximately $40\text{ }\mathrm{mW}$~\cite{zeng2008implants}. Specifically, as reported in~\cite{keppeler2018ultrafast}, optically evoked auditory responses can be successfully measured with stimulations as low as $1 \text{ }\mathrm{mW}$, while higher light-intensity will result in increased amplitude and decreased latency. In~\cite{wrobel2018optogenetic} and~\cite{keppeler2018ultrafast}, it was proven that light in the range of $470 - 640\text{ }\mathrm{nm}$ can be transmitted through optical fibers with diameter of some decades of $\mathrm{\mu m}$ and cause neural stimulation. Moreover, in~\cite{Xu2018}, the authors explored the fundamental requirements for developing a light delivery system for the cochlea and provided practical implementations. Finally, in~\cite{dombrowski2018toward}, the state-of-the-art on the emerging concept of optogenetic stimulation of the auditory pathway was presented, while the need for engineering novel multi-channel optical implants was highlighted. 

In order to deal with the limitations of the RF band, the optical wireless CI (OWCI) architecture was introduced and studied in~\cite{trevlakis2019optical}. \color{black}In the same contribution, the authors prove that OWCIs are capable of significantly improving the reliability, spectral and power efficiency by replacing the transdermal RF with an optical link. In addition, the OWCI can achieve capacity in the order of $\mathrm{Mbps}$ with only few $\mathrm{\mu W}$, in contrast to the CCI, which requires at least few $\mathrm{mW}$ in order to achieve the same capacity. Finally, the utilization of the optical frequency band provides large amounts of unexploited, non-standardized, almost-interference-free bandwidth with increased safety for the human  organism. This approach relied on several prior published works that had experimentally proven the feasibility of the transdermal optical link (TOL)~\cite{ackermann2008designing,gil2012feasibility,liu2015bidirectional}. In particular, in~\cite{ackermann2008designing}, TOLs were used to establish transdermal high-data-rate links. In~\cite{gil2012feasibility}, the key design parameters of transdermal OWC systems and their interactions were identified, accompanied by several design tradeoffs. In~\cite{liu2015bidirectional}, Liu et al. evaluated the performance of TOLs utilized for clinical neural recording purposes in terms of data-rate and transmission power. In the same work, the characteristics of the receiver were investigated with regard to its size minimization and the signal-to-noise-ratio (SNR) maximization. Despite the excellency in transmission data-rate that OWCIs provide, they cannot fully counterbalance the poor coding of spectral information in the acoustic nerve, since they carry the constraints of the electrical stimulation units. Likewise, their internal device has significant energy demands that arise from the existence of its energy-consuming digital signal processing (DSP) unit.

Based on the aforementioned research works and in combination with opto-electronic advances, several optical fiber (OF) designs were developed in order to output light in predetermined locations along the fiber body~\cite{wells2014laser,nguyen2017fabrication,pisanello2014multipoint,pisanello2018tailoring, pisano2018focused}. In more detail, in~\cite{nguyen2017fabrication}, Nguyen et al. presented a multi-point, side-firing OF capable of emitting light at multiple locations with power up to $25\%$ of the total coupled light. Furthermore, in~\cite{pisanello2014multipoint}, Pisanello et al. developed a single OF capable of delivering light to the target area through multiple target windows, while, in~\cite{pisanello2018tailoring, pisano2018focused}, the influence of several design parameters (i.e. input angle of light, numerical aperture) on the OF output light were investigated. These OFs were proposed for neural stimulation in several biomedical applications, including cochlear stimulation (see e.g.~\cite{dombrowski2018toward} and references therein).

\vspace{-0.4cm}
\subsection{Contribution}
To sum up, the hitherto proposed architectures either counterbalance the nerve stimulation limitation or the RF scarcity issue, but, none of them solves the combined problem. An architecture that would convert the audio into optical signal in the external device and transmit it directly to the cochlear nerve, would not only counterbalance the aforementioned challenges, but, it would also eliminate the necessity for an energy consuming implanted DSP device. Aspired by this, in this contribution, we propose an innovative architecture, termed as all-optical CI (AOCI), which not only exploits the characteristics of its predecessors but also is able to break their barriers. While the proposed architecture takes advantage of the beneficial particularities of the OWCI, it introduces significant modifications. The most important one is that, the AOCI is comprised only by passive components. As a result, the internal device has no power demands, thus, the proposed system is characterized by higher energy efficiency and eliminates the need of designing sophisticated power transfer approaches. Additionally, the utilization of light for the excitation of the neuron of the cochlea provides higher fidelity compared to electrical stimulations, due to the fact that light is characterized by lower spread through the human tissue. In more detail, the contribution is outlined below:
\begin{itemize}
\item We introduce the AOCI architecture and explain its building blocks (BBs) as well as their usage and functionalities.
\item We present a novel system model that accomodates the particularities of each BB, such as the light source (LS) divergence angle, the dimensions of the optical components, the channel characteristics, such as its path-gain, and the pointing errors, as well as the biological peculiarities of the human body, like the existence of neural noise. 
\item We study the AOCI feasibility and efficiency by providing the theoretical framework that quantifies its performance in terms of the average stimulation photon flux, with regard to the LS emitted optical power as well as each BBs specifications.
\item Finally, a side, yet important, contribution of this work is that we extract a novel closed-form expression for the instantaneous coupling efficiency in the presence of stochastic pointing errors.
\end{itemize}

\vspace{-0.4cm}
\subsection{Structure}
The remainder of this paper is organized as follows:  Section~\ref{S:architecture} presents the architecture of the AOCI, followed by the system model of the AOCI that is described in Section~\ref{S:sytem_model}. In Section~\ref{S:feasibility_study}, we provide the analytic framework for accessing the feasibility and efficiency of the AOCI. Respective numerical and simulation results, which illustrate the performance of the AOCI and validate the theoretical framework, alongside useful related discussions are provided in Section~\ref{S:results}. Finally, closing remarks and a summary of the main findings of this contribution are presented in Section~\ref{S:conclusions}.

\vspace{-0.4cm}
\subsection{Notations}  
Unless stated otherwise, $|\cdot|$ denotes absolute value, $\exp(\cdot)$ represents the exponential function, while $\log_{10}(\cdot)$ stands for the decadic logarithm. In addition, $P_r\left(\mathcal{A}\right)$ denotes the probability of the event $\mathcal{A}$, whereas $\mathrm{erf}(\cdot)$  denotes  the error function. Also, $J_\nu(\cdot)$ represents the Bessel function of the first kind and the $\nu$-th order, while $I_\nu(\cdot)$ represents the modified Bessel function of the first kind and the $\nu$-th order. Finally, $\Gamma(\cdot)$ is the Gamma function and $\Psi_2(\cdot)$ represents the fifth Humbert hypergeometric series~\cite[eq. 7.2.4/10]{prudnikov1988integrals}. Finally, $F^{(4)} \left[
\begin{array}{c}
\!\cdot\cdot\!:\!\cdot\!:\!\cdot\!:\!\cdot\cdot;\!\cdot;\!\cdot;\!\cdot\!:\!\cdot;\!\cdot;\!\cdot; \cdot;\!\cdot;\!\cdot\!:\!\cdot;\!\cdot;\!\cdot;\!\cdot; \\
\!\cdot\!:\!\cdot\!:\!\cdot\!:\!\cdot\cdot;\!\cdot;\!\cdot;\!\cdot\!:\!\cdot;\!\cdot;\!\cdot;\!\cdot;\!\cdot;\!\cdot\!:\!\cdot;\!\cdot;\!\cdot;\!\cdot
\end{array} 
\!\!\cdot,\!\cdot,\!\cdot,\!\cdot \right]$ is the generalized quadruple hypergeometric function~\cite[eq. 16]{saigo1988}.

\section{The AOCI Architecture} \label{S:architecture}
\begin{figure}
\centering\includegraphics[width=0.85\linewidth]{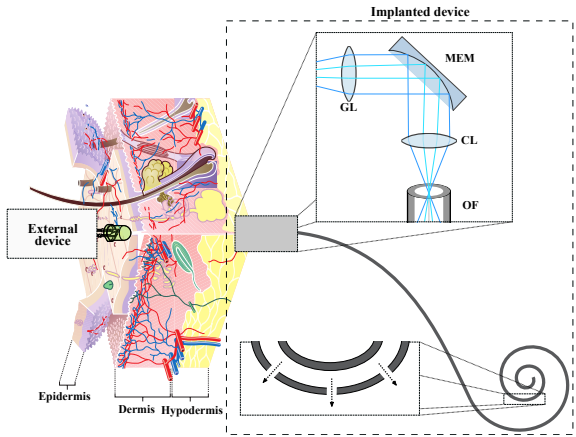}
\caption{The AOCI architecture.}\vspace{-0.5cm}
\label{architecture}
\end{figure}

As illustrated in Fig.~\ref{architecture}, the AOCI architecture consists of two devices; the external and the implanted. Much like the conventional cochlear implant, the external device is fixed on the outer surface of the skin, while the implanted one is attached on the bone of the skull, right under the skin. As a result, the total distance of the communication link is the skin thickness, which is proven to be in the range of $2-4\text{ }\mathrm{mm}$ for the area of the skull~\cite{Oltulu2018}. However, in our analysis we have considered a more pessimistic approach by considering skin thickness values between $4\text{ and }10\text{ }\mathrm{mm}$. The role of the external device is to capture the acoustic signal with a microphone, perform the necessary digital signal processing and convert it into an optical one, which is suitable for the stimulation of the acoustic nerve. Afterwards, it transmits the optical signal to the implanted device. Notice that, the main difference of the external unit of the AOCI with the corresponding CCI is that the RF front-end is replaced with a LS connected to its driver. This LS can be a light emitting diode (LED) or a laser diode (LD). The implanted unit consists of a guiding lens (GL), a microelectromechanical (MEM) device, a coupling lens (CL) and an OF. The GL is used in order to maximize the optical power captured by the implanted device via guiding the light toward the MEM, which in turn couples it into the OF via the CL. Finally, the coupled light propagates through the OF and is emitted on the photosensitive neurons of the cochlea. Next, we present in more detail the functionalities of the main BBs of the implanted device.

\underline{MEM:} Due to the unique nature of each human body (i.e. different skin thickness and color, etc.), the need arises of designing a flexible externally-operated light direction controlling system that is able to adjust AOCI to the human bodies particularities. Any failure to align the optical beam properly will result to loss of communication and, therefore, hearing loss. The solution is provided through the utilization of the MEM system, which have been greatly analyzed over the past couple of decades, due to its reduced size, light-weight and low-cost~\cite{Zou2015,Hwang2017,Zhou2017}. This system can be configured in order to not only mitigate any imperfections during the implantation process, but also customize its operation to the patient particularities. In the proposed architecture, the MEM is vital for the functionality of the system, since it enables the appropriate steering of the light beam toward the CL. In more detail, the MEM is able to adjust its optical properties by applying the appropriate electrical charge. The adjustment procedure can be performed once during the implantation process, while under stable operating state there is no need for any adjustment performed by the MEM and thus, since no energy is required, the MEM operates in a passive manner~\cite{Khoshnoud2012,Sezen2005}.

\underline{CL:} Light incident on the OF end with an angle greater than the acceptance angle will not be coupled. To prevent this, the CL is placed between the OF and the MEM to focus the signal beam in the center of the OF. Thus, the CL plays a decisive role in achieving the maximum achievable coupling efficiency. The characteristics of the CL determine the optical power that can be successfully coupled inside the OF. Note that, according to~\cite{nguyen2017fabrication}, the maximum achievable coupling efficiency cannot surpass $80\%$.

\underline{OF:} Each region of the acoustic nerve is responsible for the interpretation of a specific sound frequency in the audible spectrum. Consequently, the optical stimulations must be guided toward specific photosensitive acoustic nerve regions. This task can be accomplished by utilizing the OF architecture that was proposed in~\cite{wells2014laser}. The light propagates through the single-mode OF and the optical beam output from the OF behaves as a Gaussian beam~\cite{Marcuse1978}. It should be highlighted that, modern conventional cochlear implants house up to $22$ electrodes. However, due to the low spatial resolution of the stimulation most patients perceive sound as if only $8$ electrodes are functioning~\cite{Srinivasan2010}. In that respect, it has been proven that at least $20$ functional electrodes are needed for the patient to successfully perceive speech in a noisy environment~\cite{Friesen2001}, while this number rises to $32$ in order to recognize music~\cite{Kong2004}. Since the objective of the AOCI is to achieve superior sound perception, the number of output sites must exceed $32$. To satisfy the demand of emitting the optical signal inserted in the OF in different locations, based on its wavelength, we employ fiber Bragg gratings (FBGs), which are optical components with a periodic variation of the refractive index along the propagation direction in the core of the OF~\cite{zhou2006optic,cotillard2014regeneration}. FBGs are low-complexity structures, have low-insertion-loss, high-wavelength-selectivity and are fully compatible with single-mode OFs. Moreover, recent technological advances enabled the construction of tilted FBGs, which filter the incident light and allow only certain wavelengths to pass through, while changing the direction of the light beam according to its tilted nature~\cite{bharathan2018fiber,Mou2009}.

\vspace{-0.3cm}
\section{System model} \label{S:sytem_model}
Let us assume that the optical signal, $x$, is emitted through a transdermal wireless channel. Thus, the received signal at the implanted device can be expressed~as
\begin{align}
y_1 = h_t x,
\label{signal_y1}
\end{align}
where $h_t$ represents the optical channel coefficient~\cite{trevlakis2019optical}
\begin{align}
h_t = h_l  h_p.
\label{channel_coefficient}
\end{align}
In~\eqref{channel_coefficient}, $h_l$ and $h_p$ denote the deterministic channel coefficient, due to the propagation and the stochastic process that models the geometric spread, due to pointing error, respectively. The deterministic term of $h_t$ can be expressed~as
\begin{align}
h_l=\exp\left(-\left(\mu_\alpha+\mu_s\right) \delta\right), 
\label{pathloss}
\end{align}
where $\mu_\alpha$ represents the attenuation coefficient, $\mu_s$ is the scattering coefficient and $\delta$ is the total skin thickness. The skin attenuation and scattering coefficients can be derived from~\cite{trevlakis2018signal,Lister2012,Bosschaart2011,bashkatov2011optical,bashkatov2005optical,chan1996effects,van1989skin} and they depends on the wavelength of the optical beam. We highlight that the external device is in contact with the outer side of the skin, while the internal device is implanted on the bone of the skull. Thus, the distance between external and the internal device of the system can be approximated by $\delta$. Also, it can be observed from~\eqref{pathloss} that the signal attenuation due to scattering increases with the skin thickness. However, the pointing requirement is not affected substantially due to the limited thickness of the skin, i.e. $4-10\text{ }\mathrm{mm}$, and the necessary directivity of the optical beam captured by the implanted device.

Next, by assuming spatial intensity of beam waist, $w_{\rm \delta}$, on the implant's plane at distance $\delta$ from the LS with divergence angle, $\theta$, and circular aperture of radius $\beta$, the stochastic term of the channel coefficient can be approximated~as
\begin{align}
h_p \approx A_0\exp\left(-\tfrac{2r^2}{w_{\rm eq}^2}\right),
\label{Eq:hp1}
\end{align}
where
\vspace{-0.3cm}
\begin{align}
w_{eq}^2 &= w_{\delta}^2\tfrac{\sqrt{\pi}\erf\left(\upsilon\right)}{2\upsilon\exp\left(-\upsilon^2\right)},\\
\upsilon &= \tfrac{\sqrt{\pi}\beta}{\sqrt{2}w_{\rm \delta}},\\
w_{\rm \delta} &= \delta \tan\left(\tfrac{\theta}{2}\right), \\
A_0 &= [\erf\left(\upsilon\right)]^2.
\end{align}
Notice that $h_p$ represents the fraction of the collected power due to geometric spread with radial displacement, $r$, from the origin of the detector. Likewise, this is a well-known approximation that has been used in several reported contributions (see e.g.,~\cite{A:Outage_Capacity_for_FSO_with_pointing_errors,A:BER_performance_of_FSO_link_over_strong_atm_turbulence_channels_with_pointing_errors}, and references therein).

Moreover, it is assumed that the elevation and the horizontal displacement (sway) follow independent and identical Gaussian distributions. Hence, based on~\cite{arnon2003Effects}, it can be proven that the radial displacement at the implant follows a Rayleigh distribution with a probability density function (PDF) that can be obtained~as~\cite{B:Probability_Random_Variables_and_Stochastic_Processes}
\begin{align}
f_r\left(r\right) = \tfrac{r}{\sigma_s^2}\exp\left(-\tfrac{r^2}{2\sigma_s^2}\right), \qquad r>0 .
\label{Eq:fr1}
\end{align}
The received signal, described by~\eqref{signal_y1}, is collected by the GL and forwarded toward the MEM. The MEM output can be expressed~as
\vspace{-0.3cm}
\begin{align}
y_2 = G_c h_l h_p x ,
\label{signal_y2}
\end{align}
where $G_c$ denotes the collimation gain and, according to~\cite{sabry2017plane}, can be obtained~as 
\begin{align}
G_c = \tfrac{1}{\sqrt{\left(1-d_{in}/f\right)^2 + z_0^2/f^2}} ,
\end{align}
with $d_{in}$ being the distance between the beam waist location and the MEM surface at the point of incidence. Moreover, $f$ is the focal length of the MEM and $z_o$ is the Rayleigh range of the incident beam. In addition, the beam-waist ratio has a maximum value occurring when the input distance and the focal length are equal and can be evaluated~as~\cite{sabry2017plane}
\begin{align}
G_c^{*} = \tfrac{f}{z_0} .
\end{align}
The optical signal reflected by the MEM is captured by the CL and forwarded inside the OF. The signal coupled into the OF can be written~as
\begin{align}
y_3 = \eta G_c h_l h_p x .
\label{signal_y3}
\end{align}
where $\eta$ is the coupling efficiency. The following Theorem returns a novel closed-form expression for the evaluation of~$\eta$.

\begin{thm}
The coupling efficiency can be evaluated~as
\begin{align}
\eta\! =\! \left(\! \! \tfrac{3.83 \sqrt{2} D \omega_0}{1.22 \lambda F}\! \exp\! \! \left(\! \! -\tfrac{r^2}{\omega_0^2} \! \right)\! \Psi_2 \left(\! 1;2,1;-\tfrac{3.83^2 D^2 \omega_0^{2}}{1.22^2 \lambda^2 F^2}\!, \tfrac{r^2}{\omega_0^{2}} \! \right)\! \! \right)^2\! \! \! \!,
\label{eta_2}
\end{align}
where $\rho$, $D$, $F$ and $\omega_0$ denote the radial distance on the focal plane, the focusing lens diameter, focal length and the OF mode field radius, respectively.
\end{thm}
\begin{IEEEproof}
Please refer to Appendix A. 
\end{IEEEproof}
From~\eqref{eta_2}, we observe that $\eta$ depends on the diameter and focal length of the CL, the OF mode field radius, the transmission wavelength and the pointing errors. 

The signal coupled into the OF is subject to bending and FBG losses. Thus, the optical signal that is emitted on the acoustic nerve, can be written~as
\begin{align}
y_4 = k \eta G_c h_l h_p x ,
\label{signal_y4}
\end{align}
where $k$ is the light beam's propagation efficiency through the OF. The bending losses of conventional OFs can be calculated using weakly guiding or adiabatic approximation~\cite{yu2009modeling}. However, these approximation are not valid for sharply bent microfiber, which are usually high-index-contrast waveguides. In practice, the strong optical confinement ability of the microfiber limits the power leakage ($0.14\text{ }\mathrm{dB/90^{\circ}}$) even for increased bending radius or index values. Also, the losses generated by the use of FBGs are proven to be in the order of $10\%$~\cite{Wang2016}.

The stimulation of the cochlear neurons depends on the number of photons emitted at the output of the OF~\cite{Hadfield2009}. When the received photons surpass a threshold, the neurons are excited and an auditory brainstem response (ABR) is triggered. In order to determine the amount of photons, we convert the optical power of the optical signal that is emitted on the cochlear neurons to the corresponding photon flux, $\Phi$, which can be expressed~as~\cite{Hadfield2009}
\begin{align}
\Phi = \tfrac{\lambda}{h c} y_4,
\end{align}
where $h$ and $c$ denote the Plank's constant and the speed of light, respectively.

As indicated by the above analysis, the AOCI architecture consists of only passive components; therefore, no-noise is generated in the signal at the output of the OF. However, given the photosensitive nature of the cochlear neurons and the fact that a background fluorescence exists in the human body, a photon-shot-noise is generated~\cite{Sjulson2007,Hendel2008,Sasaki2008,Vogelstein2009,wilt2013photon}. In more detail, the background fluorescence that exists inside the human body is generated by unlabeled cellular elements, as well as properly and improperly targeted indicators received by the neurons. In addition, it is represented by the rate of detected photons, $F_0$, in the absence of a transmission and combines fluorescence excitation, emission, and detection. The shot-noise arises from the quantum nature of light and follows a Poisson distribution, since photons arrive at the neuron continuously and independently. As a result, the additive noise follows a Poisson distribution with mean $\overline{B} = F_0 \tau$, where $\tau$ is the neurons time decay constant. The probability mass function (pmf) of the photon-shot-noise can be obtained~as~\cite{wilt2013photon}
\begin{align}
p_N(n) = \tfrac{\overline{B}^n}{n!} e^{-\overline{B}} ,
\end{align}
where $n$ are the photons generated by the background fluorescence. As a result, the photon flux signal model can be written~as
\vspace{-0.3cm}
\begin{align}
Y = \Phi + N ,
\end{align}
where $Y$ and $N$ represent the photon flux received by the neurons and the background photon-shot-noise, respectively.

\section{A Feasibility Study} \label{S:feasibility_study}
In order to validate the feasibility of the AOCI and reveal the appropriate values of the design parameters, we derive the average photon flux.

\begin{thm}
The average emitted photon flux can be evaluated as in \eqref{mean_phi_2}, given at the top of the next page.
\end{thm}
\begin{IEEEproof}
Please refer to Appendix B. 
\end{IEEEproof}

\begin{figure*}
\begin{flalign}
\begin{aligned}
\overline{\Phi} = \! \tfrac{k G_c h_l \lambda x A_0 \left(\! \tfrac{3.83 \sqrt{2} D \omega_0}{1.22 \lambda F\!} \right)^2}{2 h c \sigma_s^2 \left( \tfrac{2}{\omega_0^2}\! +\tfrac{2}{w_{eq}^2}+\! \tfrac{1}{2\sigma_s^2} \right)} F^{(4)} \left[
\begin{array}{c}
\!\!\text{-}\!\!::: \!\!\text{-}; \!\!\text{-}; \!\!\text{-}; \!\!\text{-}: \!\!\text{-}; 1; \!\!\text{-}; 1; \!\!\text{-}; 1: \!\!\text{-}; \!\!\text{-}; \!\!\text{-}; \!\!\text{-}; \\
\!\!\text{-}\!\!::: \!\!\text{-}; \!\!\text{-}; \!\!\text{-}; \!\!\text{-}: \!\!\text{-}; \!\!\text{-}; \!\!\text{-}; \!\!\text{-}; \!\!\text{-}; \!\!\text{-}: 2;  2;  1;  1;
\end{array} 
\!\!\! \text{-}\tfrac{3.83^2 D^2 \omega_0^{2}}{1.22^2 \lambda^2 F^2},\!\!\text{-}\tfrac{3.83^2 D^2 \omega_0^{2}}{1.22^2 \lambda^2 F^2},\!\!\tfrac{\tfrac{1}{\omega_0^{2}}}{\tfrac{2}{\!\omega_0^2}\! +\!\tfrac{2}{w_{eq}^2}\!+\! \tfrac{1}{2\sigma_s^2}\!},\!\!\tfrac{\tfrac{1}{\omega_0^{2}}}{\tfrac{2}{\!\omega_0^2}\! +\!\tfrac{2}{w_{eq}^2}\!+\! \tfrac{1}{2\sigma_s^2}\!}\! \right] .
\label{mean_phi_2}
\end{aligned}
\end{flalign}
\line(1,0){512}
\end{figure*}

From~\eqref{mean_phi_2} it becomes evident that the average emitted photon flux is dependent upon three terms, i.e. $\left(x A_0 e^{-(\mu_\alpha+\mu_s)\delta}/\lambda \sigma_s^2 \right)$, $\left( D \omega_0 / F \right)^2$ and $\left( 2 / \omega_0^2 + 2 / w_{eq}^2 + 1/ 2 \sigma_s^2 \right)^{-1}$. The first one models the impact of the LS and the transdermal channel. We observe the average emitted photon flux increases with the transmission power, while it decreases with the wavelength, the skin thickness and the misalignment standard deviation. The second term contains the transmission wavelength and the design parameters related to the coupling of the light, i.e. OF mode field radius, CL focal length and diameter. This term is proportional to $\overline{\Phi}$ and has a more detrimental effect due to the fact that it is raised to the square. However, it should be noted that the selection of the CL and OF dimensions is bound by the geometry of the coupling phenomenon. In other words, the selection of a higher diameter CL does not ensure a higher emitted photon flux. Finally, due to the last term, $\overline{\Phi}$ increases with the OF mode field radius, the misalignment standard deviation and the equivalent beam radius.

However, after the successful excitation of a neuron, it enters a relaxation period of duration $\tau$.
This behavior can be modeled based on~\cite{wilt2013photon} and the received photon flux can be evaluated~as 
\vspace{-0.3cm}
\begin{align}
Y_1 = \overline{\Phi} \exp{\left(-\tfrac{t}{\tau}\right)} + F_0 .
\end{align}
In addition, when a signal is transmitted, the average photon flux of the signal plus the background noise is their time integral, which can be expressed~as
\begin{align}
\overline{Y_1} = \int_0^\tau Y_1 dt = \overline{\Phi} \tfrac{\tau\left(e-1\right)}{e} + \overline{B} .
\label{Y1}
\end{align}
Notice that~\eqref{Y1} is the link budget of the proposed architecture and that the average received photon flux is proportional to the average emitted photon flux and the neural relaxation period. This indicates that the link budget is affected by the characteristics of the optical devices, the transdermal channel path-loss and pointing errors, as well as the background photon-shot-noise.

Although, $\overline{\Phi}$ is a widely-accepted feasibility study metric, due to the stochastic nature of the channel, it is unable to fully quantify the performance of the AOCI. Motivated by this, we defined three key-performance-indicators (KPIs), which assess the successful neural stimulation (probability of hearing), the unintentional neural stimulation (probability of false-hearing) and the neural damage (probability of neural damage). To the best of the authors knowledge, this work is one of the first that provides an engineering perspective to the important topic of CIs. As a consequence, there are no previously published papers that define these KPIs.

\textbf{Definition 1.}
The probability of hearing can be defined~as
\begin{align}
P_h = P_r(Y\geq\mathcal{Y}_{th}) ,
\end{align}
where $\mathcal{Y}_{th}$ is the minimum required photon flux in order to achieve neural excitation. 

\textbf{Definition 2.}
The probability of false-hearing is definde as the probability to excite a neuron in the absence of transmitted signal and can be evaluated~as
\begin{align}
P_m = P_r(N\geq\mathcal{Y}_{th}) .
\label{Pm1}
\end{align}

The following theorem returns a closed-form expression for the probability of false-hearing.
\begin{thm}
The false-hearing probability can be~evaluated~as
\begin{align}
P_m = \tfrac{\Gamma(\mathcal{Y}_{th}+1,\overline{B})}{\mathcal{Y}_{th}!} .
\label{Pm3}
\end{align}
\end{thm}
\begin{IEEEproof}
From~\eqref{Pm1}, we can rewrite the false-hearing probability~as 
\vspace{-0.3cm}
\begin{align}
P_m = \mathcal{F}_N(\mathcal{Y}_{th}) ,
\label{Pm2}
\end{align}
where $\mathcal{F}_N$ is the cumulative density function (CDF) of $N$. Since, N is a Poisson distributed random variable, its CDF can be obtained~as 
\begin{align}
\mathcal{F}_N(\mathcal{Y}_{th}) = \tfrac{\Gamma(\mathcal{Y}_{th}+1,\overline{B})}{\mathcal{Y}_{th}!} .
\label{N_CDF}
\end{align}
From~\eqref{Pm2} and~\eqref{N_CDF}, we get~\eqref{Pm3}. This concludes the proof.
\end{IEEEproof}

\textbf{Definition 3.}
The probability of neural damage is defined~as
\begin{align}
P_d = P_r(N\geq\mathcal{D}_{th}|\xi=0) + P_r(Y\geq\mathcal{D}_{th}|\xi=1) ,
\label{Pd1}
\end{align}
where $P_r(N\geq\mathcal{D}_{th}|\xi=0)$ and $P_r(Y\geq\mathcal{D}_{th}|\xi=1)$ are the conditional probabilities that the instantaneous photon flux surpasses the the maximum permissible exposure (MPE) for the cochlear neurons, $\mathcal{D}_{th}$, in the absence and presence of optical signal, respectively. Additionally, $\xi\mathcal{\epsilon}\{0,1\}$ is a binary variable that characterizes the absence ($\xi=0$) and presence ($\xi=1$) of optical signal. 

Notice that, in practice, $N\ll\mathcal{D}_{th}$, hence,~\eqref{Pd1} can be approximated as
\begin{align}
P_d \approx P_r(Y\geq\mathcal{D}_{th}) .
\label{Pd2}
\end{align}

\section{Numerical Results and Discussions} \label{S:results}
In this section, we report numerical results, accompanied by related discussions, which highlight the feasibility and efficiency of the AOCI. In addition, insightful design guidelines for the BBs of the AOCI are extracted. Additionally, we demonstrate and compare the theoretical results with Monte Carlo simulations that verify the analytic framework. In the following figures, the dashed-line specifies the minimum photon flux required for the excitation of the cochlear neurons~\cite{wrobel2018optogenetic}. In addition, we point out that based on~\cite{wrobel2018optogenetic}, the highest LS optical transmission power, $200\text{ }\mathrm{mW}$, corresponds to irradiance of approximately $12\text{ }\mathrm{mW/mm^2}$ on the cochlear neurons and $56\text{ }\mathrm{mW/mm^2}$ on the skin, which are below the respective MPE limits ($\sim75\text{ }\mathrm{mW/mm^2}$ and $500\text{ }\mathrm{mW/mm^2}$)~\cite{Han2012,ICNIRP2013a}. Finally, the simulation parameters are presented in Table~\ref{tab:parameters} alongside the corresponding symbols, values and references.

\vspace{-0.2cm}
\begin{table}[h]
\centering
\caption{Simulation parameters.}
\label{tab:parameters}
\begin{tabular}{l l l l}
\hline
Parameter & Symbol & Value & References \\
\hline
LS divergence angle & $\theta$ & $[5-30]^{\circ}$ & \cite{ackermann2008designing} \\
Wavelength & $\lambda$ & $594\text{ }\mathrm{nm}$ & \cite{keppeler2018ultrafast,Moser2020} \\
Skin thickness & $\delta$ & $[4-10]\text{ }\mathrm{mm}$ & \cite{liu2015bidirectional}\\
CL radius & $D$ & $0.1\text{ }\mathrm{mm}$ & \cite{Jahns2008,Yang2004} \\
OF mode field radius & $\omega_0$ & $0.1\text{ }\mathrm{mm}$ & \cite{wrobel2018optogenetic,Moser2020}\\
Time decay constant & $\tau$ & $0.15\text{ }\mathrm{s}$ & \cite{wilt2013photon} \\
\hline
\end{tabular}
\end{table}

\begin{figure}
\centering\includegraphics[width=0.73\linewidth]{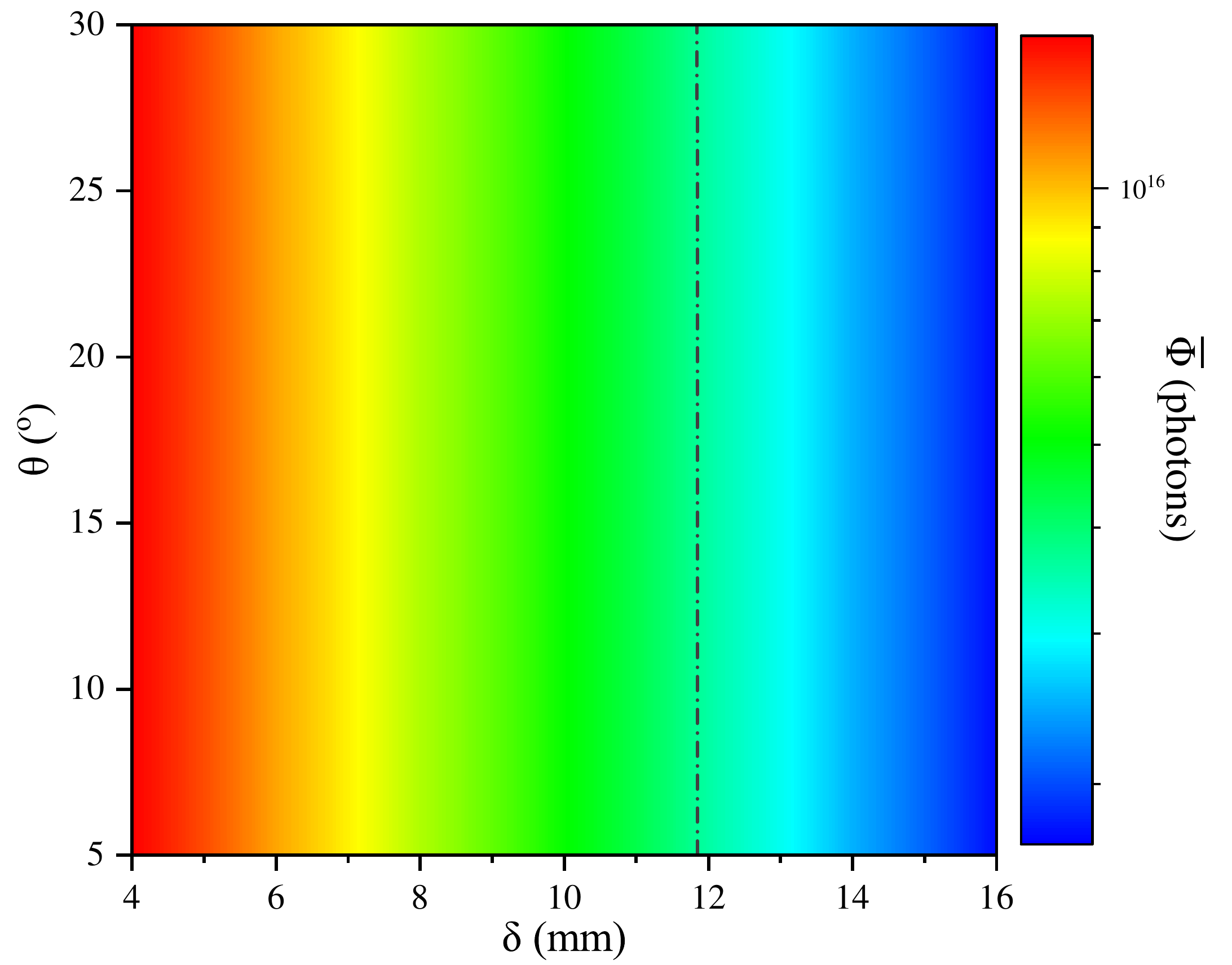}
\caption{Average emitted photon flux vs skin thickness and LS divergence angle, for optical transmission power of $20\text{ }\mathrm{mW}$.} \vspace{-0.4cm}
\label{phi_vs_theta_and_delta}
\end{figure}
In Fig.~\ref{phi_vs_theta_and_delta}, the average photon flux is depicted as a function of the skin thickness and the divergence angle for optical transmission power of $20\text{ }\mathrm{mW}$. From this figure, it is evident that for a fixed divergence angle, the average emitted photon flux increases as the skin thickness decreases. For example, for divergence angle equal to $20^{\circ}$, the average emitted photon flux increases by approximately $44\%$, as the skin thickness decreases from $8$ to $6\text{ }\mathrm{mm}$. This is expected because as the transmission distance increases, the amount of the optical power captured by the implanted device decreases. Additionally, a decrease in divergence angle, for the same skin thickness, results in an increase in the emitted photon flux. For instance, for a fixed skin thickness equal to $6\text{ }\mathrm{mm}$, as the divergence angle decreases from $30^{\circ}$ to $20^{\circ}$, an $3\%$ increase of the emitted photons is observed. This increase is anticipated because, for higher divergence angle values, the amount of light guided far from the implanted device increases. Finally, we highlight the fact that the path-loss has a more detrimental effect on the AOCI performance, compared to the non-concentration of the optical beam. 

\begin{figure}
\centering\includegraphics[width=0.75\linewidth]{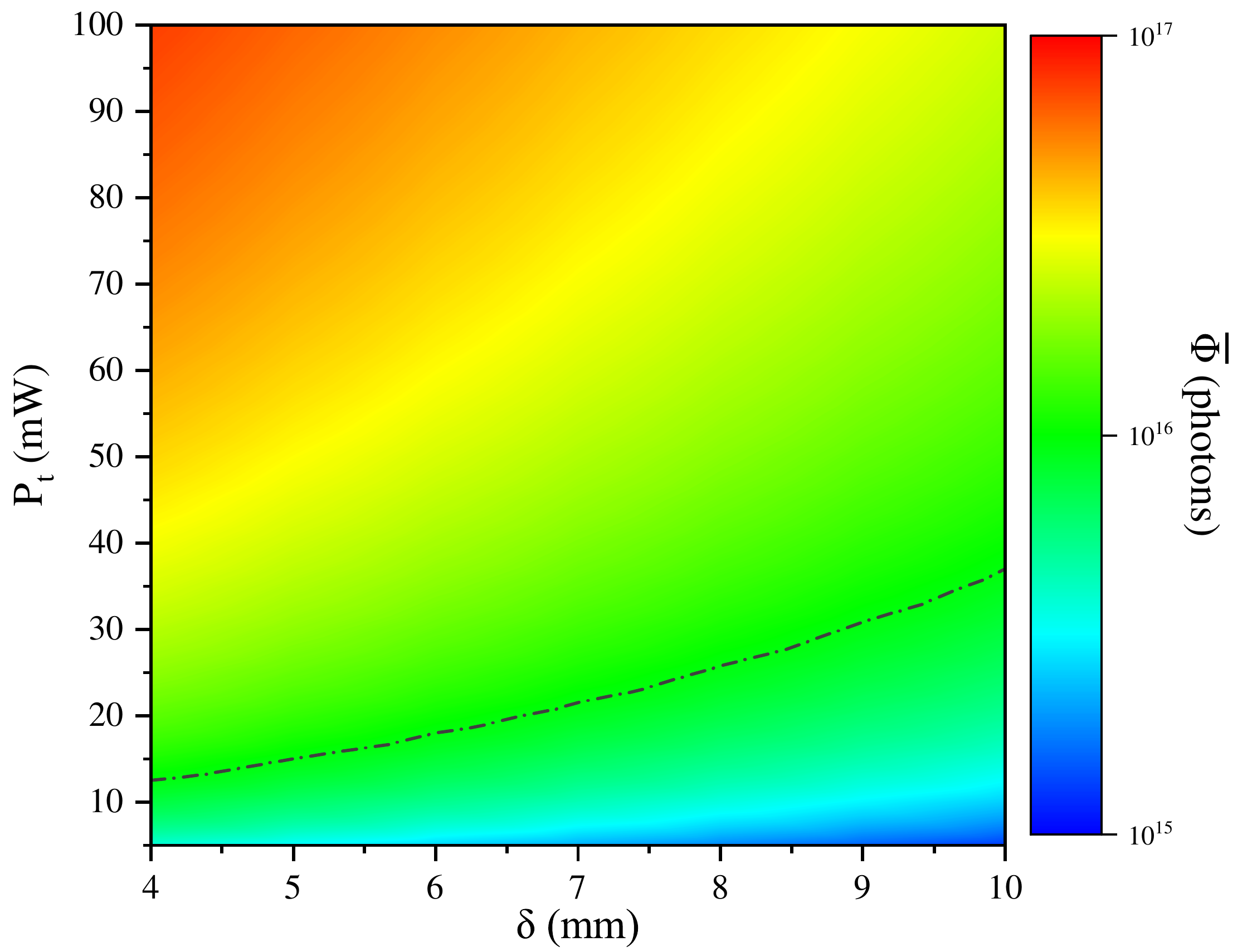}
\caption{Average emitted photon flux vs skin thickness and optical transmission power.} \vspace{-0.4cm}
\label{phi_vs_delta_and_p}
\end{figure}
In Fig.~\ref{phi_vs_delta_and_p}, the average emitted photon flux is presented as a function of the skin thickness and optical transmission power. From this figure, it becomes obvious that, for a fixed skin thickness, as the optical transmission power of the LS decreases, the average emitted photon flux also decreases. For instance, for a skin thickness set to $6\text{ }\mathrm{mm}$, as the optical transmission power decreases from $20$ to $10\text{ }\mathrm{mW}$, the average emitted photon flux decreases approximately $49\%$. Meanwhile, for the same optical transmission power as the skin thickness increases, the number of the emitted photons decreases. Likewise, this figure provides an illustration of the equilibrium formed between the gain from increasing optical transmission power and the possible restrictions imposed on the AOCI from the skin thickness. In other words, it indicates that for patients with increased skin thickness, the AOCI should use a higher power to achieve the same performance, in terms of the average emitted photon flux.

\begin{figure}
\centering\includegraphics[width=0.71\linewidth]{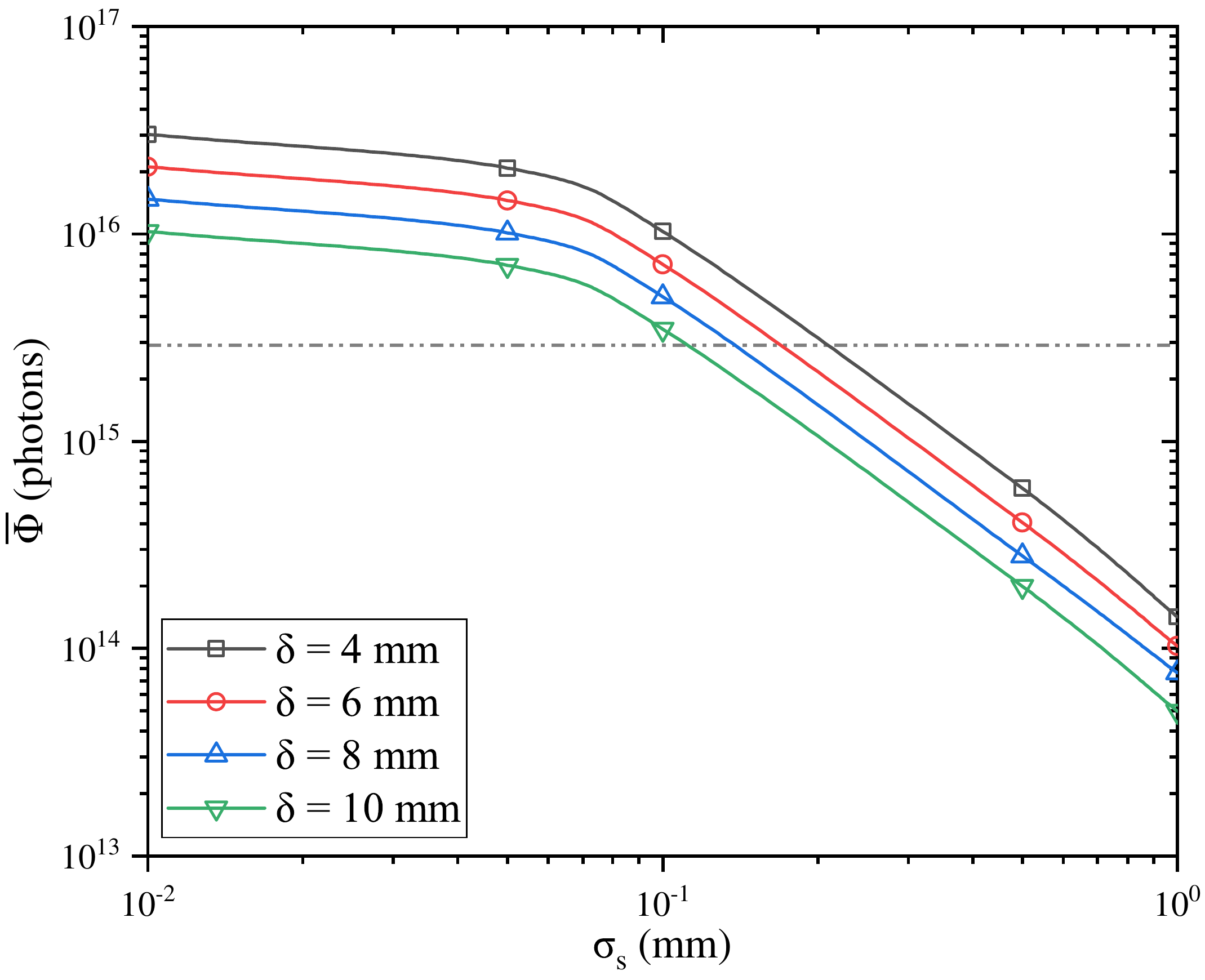}
\caption{Average emitted photon flux vs misalignment standard deviation, for different values of skin thickness, and optical transmission power of $40\text{ }\mathrm{mW}$.} \vspace{-0.4cm}
\label{phi_vs_jitter_and_delta}
\end{figure}
In Fig.~\ref{phi_vs_jitter_and_delta}, the average emitted photon flux is illustrated as a function of the misalignment standard deviation, for different values of skin thickness and optical transmission power of $40\text{ }\mathrm{mW}$. It is obvious that simulation and analytic results coincide, which verifies the validity of the theoretical framework. From this figure, we observe that for a certain standard deviation of the pointing error, as the skin thickness increases, the average emitted photon flux decreases. This is expected because the transmission distance of the optical signal increases; thus, the channel attenuation is higher. Moreover, for a fixed skin thickness, the average emitted photon flux is inversely proportional to the pointing error standard deviation. For instance, for skin thickness equal to $6\text{ }\mathrm{mm}$, as the pointing error standard deviation varies from $0.1$ to $1\text{ }\mathrm{mm}$, the average emitted photon flux decreases by approximately $98\%$, i.e. from $7.13 \times 10^{15}$ to $1.03 \times 10^{14}$ photons. It is important to highlight that the misalignment between the LS and the implanted device is the determining factor.

\begin{figure}
\centering\includegraphics[width=0.71\linewidth]{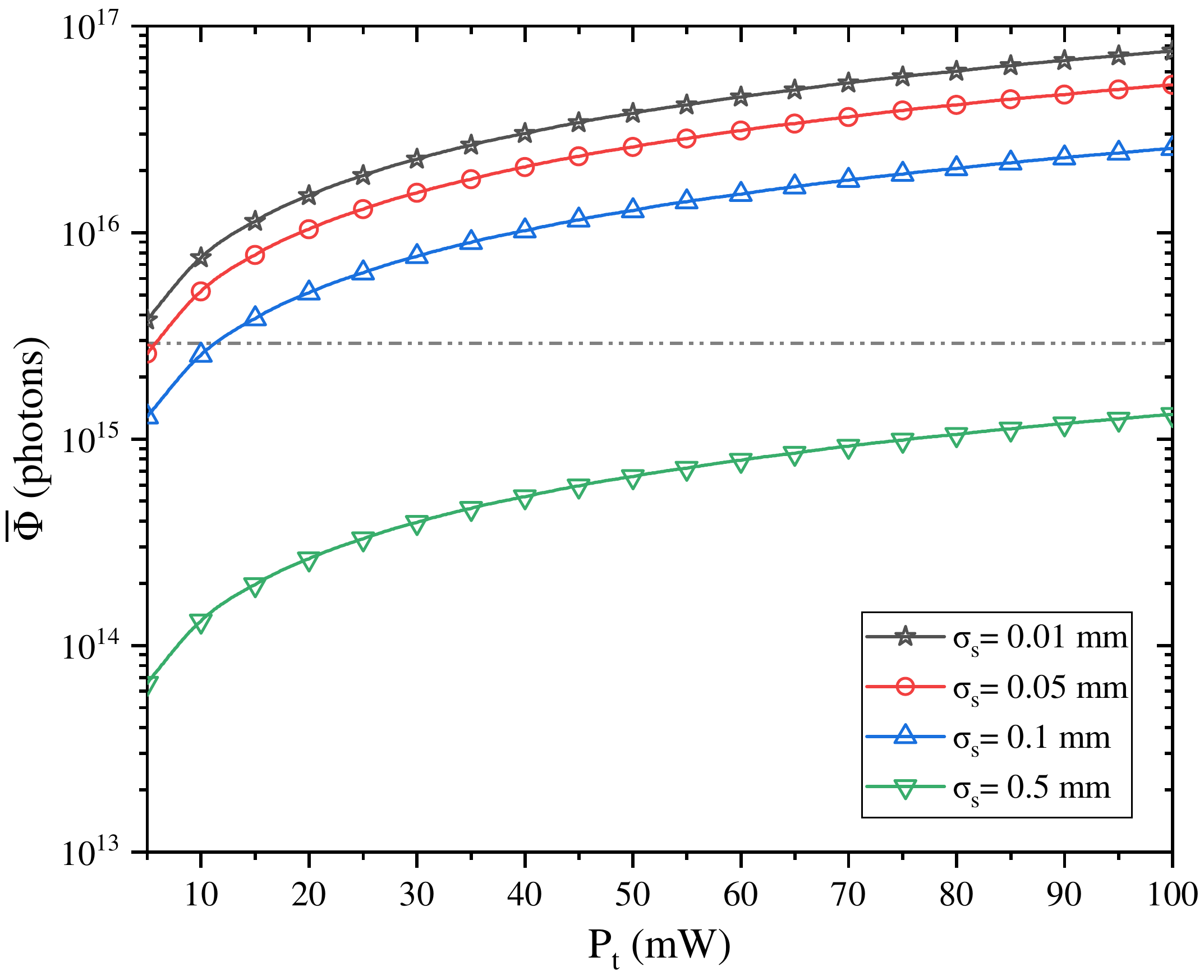}
\caption{Average emitted photon flux vs optical transmission power, for different values of misalignment standard deviation.} \vspace{-0.4cm}
\label{phi_vs_p_and_jitter} 
\end{figure}
In Fig.~\ref{phi_vs_p_and_jitter}, the average emitted photon flux is illustrated as a function of the optical transmission power, for different values of the pointing errors standard deviation. As expected, for a fixed pointing error standard deviation value, an increase in the optical transmission power, results in increased number of emitted photons on the cochlear neurons. For example, for a pointing errors standard deviation equal to $0.1\text{ }\mathrm{mm}$, as the optical transmission power increases from $10$ to $20\text{ }\mathrm{mW}$, the average emitted photon flux clearly surpasses the excitation limit of the cochlear neurons. This is of high significance for the design of the AOCI because, on the one hand, the effect of the pointing errors can be mitigated by increasing $P_t$ without surpassing the safe limits, while on the other hand, the proposed architecture is proven capable of overcoming the limitations that are entangled with the uniqueness of each human organism. Additionally, it is worth-noting that, for a fixed optical transmission power, an increase in the pointing errors standard deviation results in a decrease of the average emitted photon flux. For instance, for $20\text{ }\mathrm{mW}$ optical transmission power, as the pointing errors standard deviation decreases from $0.5$ to $0.1\text{ }\mathrm{mm}$, the average emitted photon flux increases by about $94\%$. However, for the same optical transmission power, as the pointing errors standard deviation decreases from $0.05$ to $0.01\text{ }\mathrm{mm}$, the average emitted photon flux increases by $45\%$. This fact highlights once more the detrimental effect of the pointing errors on the performance of the AOCI and the importance to mitigate it.  

\begin{figure}
\centering\includegraphics[width=0.71\linewidth]{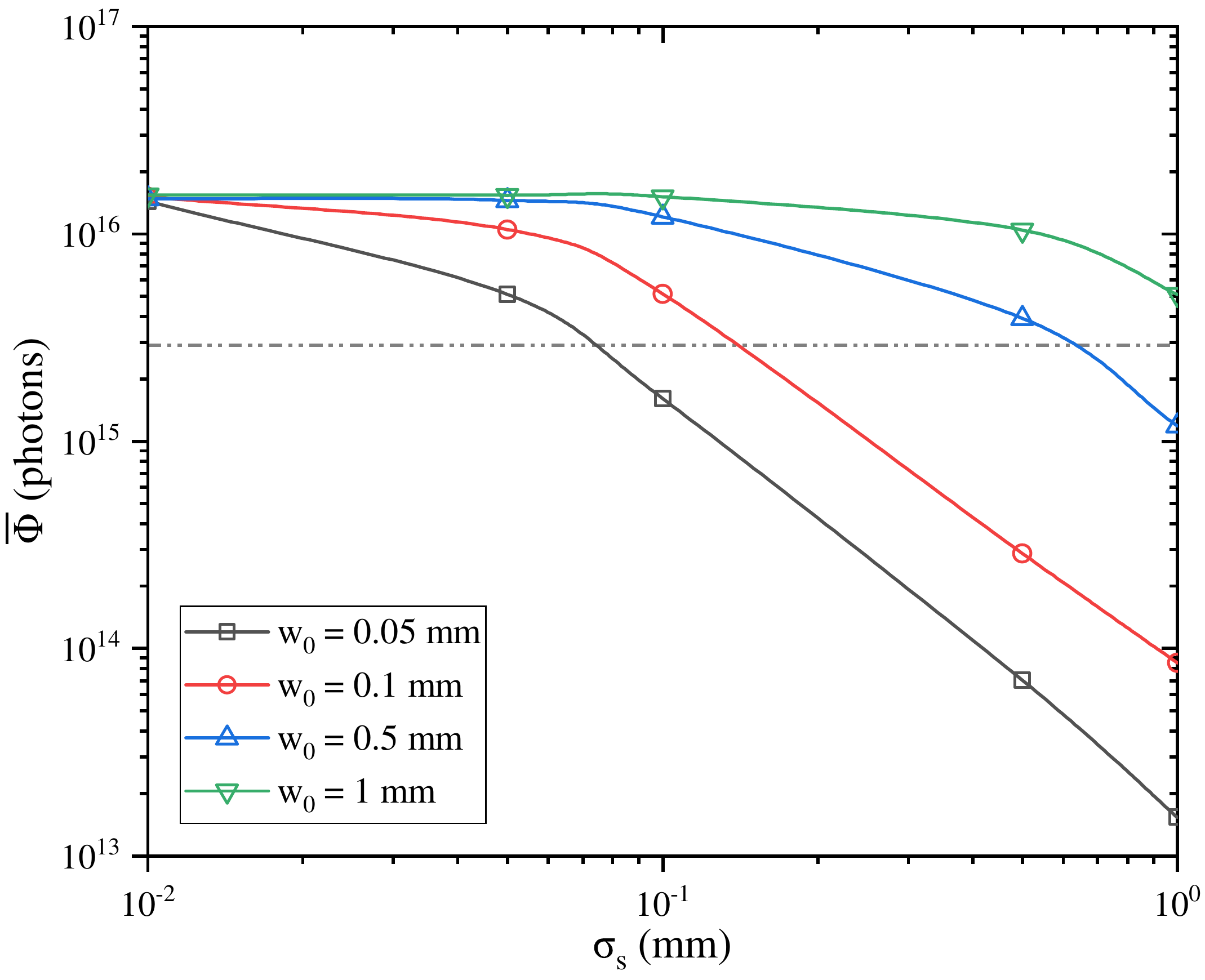}
\caption{Average emitted photon flux vs misalignment standard deviation, for different OF dimensions, and optical transmission power of $20\text{ }\mathrm{mW}$.} \vspace{-0.4cm}
\label{phi_vs_jitter_and_w0}
\end{figure}
A possible solution for the mitigating the impact of the pointing errors is provided by altering the characteristics of the CL and the OF, as depicted in Fig.~\ref{phi_vs_jitter_and_w0}. In particular, in Fig.~\ref{phi_vs_jitter_and_w0}, the average emitted photon flux is illustrated as a function of the pointing errors standard deviation, for different CL and OF dimensions, and optical transmission power of $20\text{ }\mathrm{mW}$. The validity of the theoretical framework is verified through the coincidence of the simulation and analytic results. We observe that, for a specific CL and OF pair, as the pointing errors standard deviation increases, the emitted average emitted photon flux decreases. However, for a fixed value of pointing errors standard deviation, different characteristics of the CL and OF can achieve different average emitted photon flux performance. It is highlighted that, the selected values for CLs and OFs, which are employed in this figure, have been carefully selected in order to maintain the highest achievable coupling efficiency, i.e. approximately $80\%$. In more detail, as the pointing errors standard deviation increases, the impact of the characteristics of the CL and the OF on the performance of the system increases, as well. For instance, for mode field radius of $0.1$ and $1\text{ }\mathrm{mm}$ and the corresponding characteristics for the CL, the average emitted photon flux degrades by approximately $99\%$ and $67\%$, respectively, as the pointing errors standard deviation varies from $0.1$ to $1\text{ }\mathrm{mm}$. This observation reveals that it is possible to mitigate the impact of the pointing errors on the AOCI's performance by increasing the mode field radius of the OF.

\begin{figure}
\centering\includegraphics[width=0.72\linewidth]{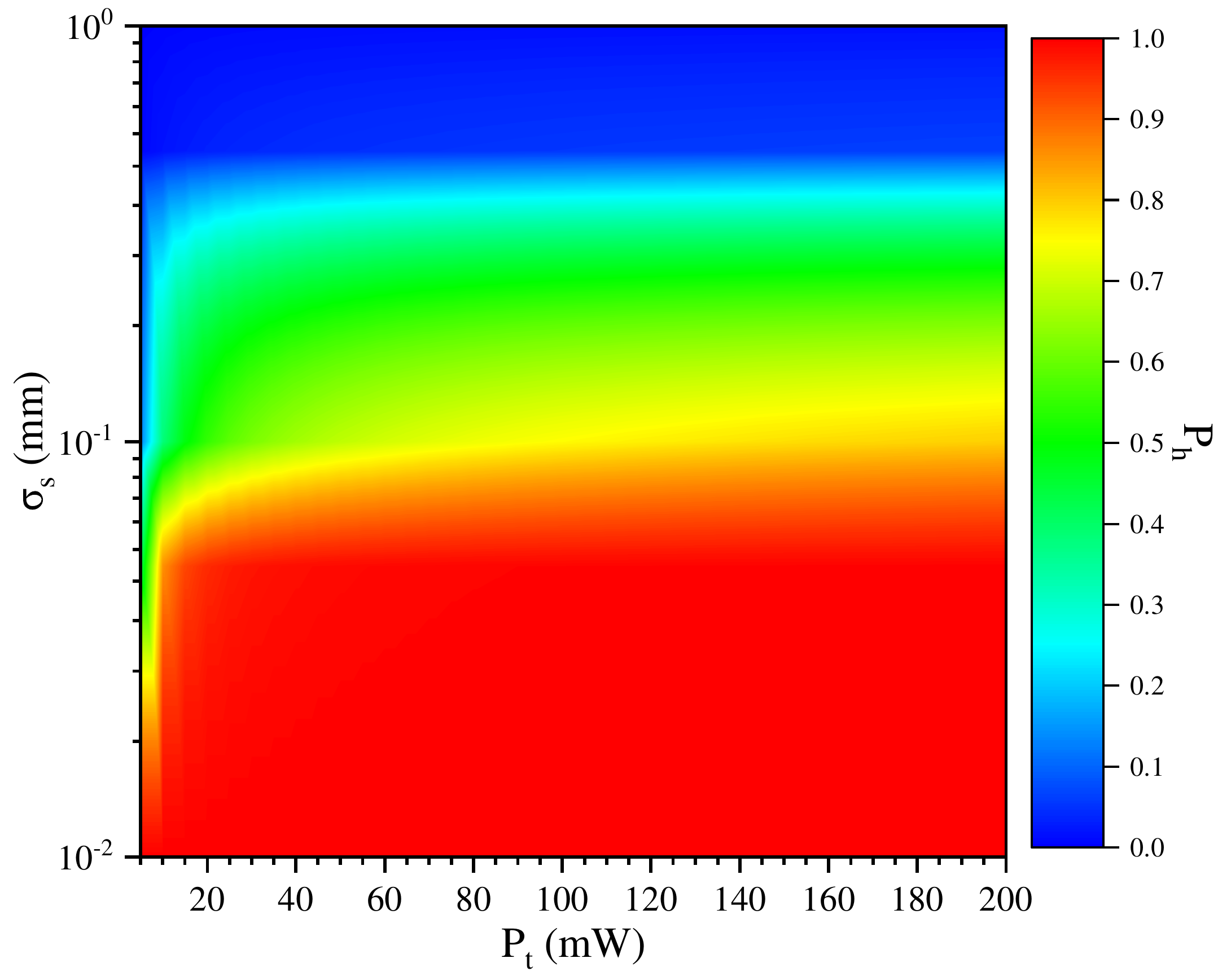}
\caption{Probability of hearing vs misalignment standard deviation and optical transmission power.} \vspace{-0.4cm}
\label{Ph_vs_p_and_jitter}
\end{figure}
In Fig.~\ref{Ph_vs_p_and_jitter}, the probability of hearing is depicted as a function of the pointing errors standard deviation and the optical transmission power. As expected, for a given pointing errors standard deviation, the probability of hearing increases proportionately to the optical transmission power. For example, for pointing errors standard deviation of $10^{-1}\text{ }\mathrm{mm}$, as the optical transmission power varies from $20$ to $120\text{ }\mathrm{mW}$, the probability of hearing increased by about $36\%$. In addition, for a fixed value of the optical transmission power, as the pointing errors standard deviation decreases the probability of hearing increases. For instance, for a optical transmission power of $100\text{ }\mathrm{mW}$, as the pointing errors standard deviation increases from $0.1$ to $0.5\text{ }\mathrm{mm}$, the probability of hearing decreases by approximately $93\%$. These observations highlight the detrimental impact of the pointing errors on the systems performance, which can me mitigated by altering the optical components characteristics and increasing the optical transmission power. Finally, from this figure it becomes noticeable that for realistic values of the pointing errors standard deviation (i.e. $\sigma_s \leq 0.05\text{ }\mathrm{mm}$), the AOCI achieves a probability of hearing higher than $90\%$ with an optical transmission power in the order of $20\text{ }\mathrm{mW}$, while, according to~\eqref{Pm3}, the probability of false-hearing is approximately zero. Note that CCIs demand a significantly additional transmission power in order to cover the needs of the implanted device, while the implanted device of the AOCI is composed only from passive elements. This indicates that the AOCI approach is a greener biomedical device paradigm in comparison with the conventional ones.

\begin{figure}
\centering\includegraphics[width=0.71\linewidth]{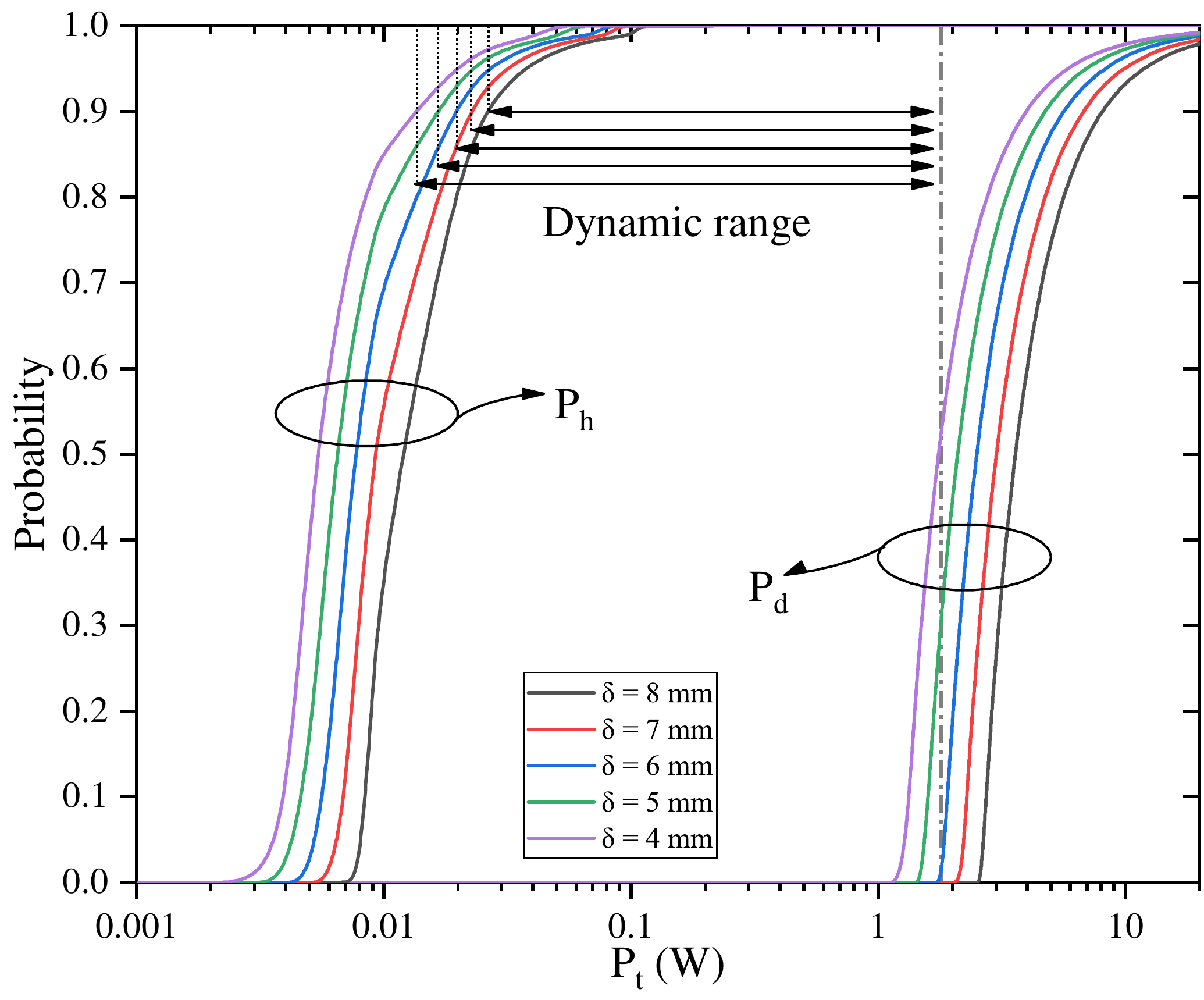}
\caption{Probability of hearing and probability of neural damage vs optical transmission power for different values of skin thickness, and misalignment standard deviation of $0.05\text{ }\mathrm{mm}$.} \vspace{-0.4cm}
\label{CombinedP_vs_p_and_delta}
\end{figure}
Fig.~\ref{CombinedP_vs_p_and_delta} presents the probability of hearing and the probability of neural damage as a function of the optical transmission power for different skin thickness values. The dashed-line specifies the MPE limit for the optical optical beam irradiance incident on the skin~\cite{ICNIRP2013a}. Note that, the values of the optical transmission power on the horizontal axis are presented in a logarithmic scale. We observe that for a specific skin thickness, the probability of neural damage increases with the optical transmission power, which is expected. Moreover, from this figure it becomes evident that for a specific value of optical transmission power the acoustic nerve damage probability is highly influenced by the skin thickness. For instance, for optical transmission power of $2\text{ }\mathrm{W}$ and skin thickness lower than $7\text{ }\mathrm{mm}$, the damage of acoustic nerve is almost certain. It is highlighted that the values of the optical transmission power, for which the probability of neural damage is defined, i.e. higher than $1\text{ }\mathrm{W}$, are extremely high. Moreover, the MPE limit for the transdermal link is more strict than the probability of neural damage; thus, it defines the safe dynamic range of the AOCI system. In more detail, the dynamic range of the AOCI is inversely proportional to the skin thickness, i.e. as the skin thickness increases, the dynamic range decreases. For example, as skin thickness varies from $8$ to $4\text{ }\mathrm{mm}$, the dynamic range increases approximately $2\%$. Finally, from this figure it is apparent that the proposed AOCI architecture is safe for optogenetic applications.

\vspace{-0.3cm}
\section{Conclusions} \label{S:conclusions}
This paper was devoted on delivering an innovative AOCI architecture and describing its BBs. Moreover, a system model that incorporates its particularities, a link budget analysis that theoretically verifies its feasibility, as well as the probabilities of hearing, false-hearing and neural damage, which quantify its efficiency and safety for the human body, are presented. The accuracy of the theoretical analysis was verified through Monte Carlo simulations, which additionally revealed that the AOCI outperforms CCIs in terms of energy consumption, and constitutes a safe solution for a significant number of patients. Finally, both the system model and theoretical framework are expected to become the basic design tools of AOCIs. 

\vspace{-0.2cm}
\section*{Appendices}
\vspace{-0.1cm}
\section*{Appendix A}
\label{Appendix_eta}
If we assume that a Gaussian optical beam is focused on a single-mode fiber core, the coupling efficiency is given by~\cite[eq. 6]{Toyoshima2006}
\begin{align}
\eta = \tfrac{\left| \int_0^\infty \! A(\rho)M(\rho)d\rho \right|^2}{\int_0^\infty \! \left| A(\rho) \right|^2 d\rho} ,
\label{eta}
\end{align}
where, $A(\rho)$ and $M(\rho)$ can be expressed as~\cite[eq. 2]{Toyoshima2006}
\begin{align}
A(\rho) = \tfrac{\pi D^2}{2 \lambda F} \tfrac{J_1 \left(\tfrac{3.14 \rho D}{\lambda F}\right)}{\left(\tfrac{3.14 \rho D}{\lambda F}\right)} ,
\label{A}
\end{align}
and 
\vspace{-0.4cm}
\begin{align}
M(\rho) = \tfrac{2\sqrt{2 \pi} \rho}{\omega_0} \exp\left(-\tfrac{\rho^2 + r^2}{\omega_0^2}\right) I_0\left(\tfrac{2 \rho r}{\omega_0^2}\right) .
\label{M}
\end{align}

By substituting \eqref{A} and \eqref{M} into \eqref{eta}, the coupling efficiency can be rewritten as follows
\begin{align}
\eta\! =\! \left|\! \tfrac{2 \sqrt{2}}{\omega_0}\! \right|^2\! \left|\! \int_0^\infty \! \! \! \! J_1\! \left(\! \tfrac{3.14 \rho D}{\lambda F}\! \right)\! \exp\! \left(\! - \tfrac{\rho^2\! +\! r^2}{\omega_0^2}\! \right)\! I_0\!\left(\! \tfrac{2 \rho r}{\omega_0^2}\! \right)\! d\rho \right|^2 .
\label{eta_1}
\end{align}
The coupling efficiency, from~\eqref{eta_1}, can be written as follows
\begin{align}
\eta\! =\! \left|\! \tfrac{2 \sqrt{2}}{\omega_0}\! \exp\! \left(\! -\tfrac{r^2}{\omega_0^2}\right)\! \right|^2\! \left| \mathcal{I} \right|^2 ,
\label{eta_3}
\end{align}
where
\vspace{-0.4cm}
\begin{align}
\mathcal{I} = \int_0^\infty \! J_1\! \left(\! \tfrac{3.83 \times 2 \rho D}{1.22 \lambda F}\! \right)\! \exp\! \left(\! -\tfrac{\rho^2}{\omega_0^2}\right)\! I_0\!\left(\! \tfrac{2 \rho r}{\omega_0^2}\! \right)\! d\rho .
\label{Ia}
\end{align}
By employing~\cite[eq. 8.441/2]{gradshteyn2014table} and~\cite[eq. 8.447/1]{gradshteyn2014table}, \eqref{Ia} can be rewritten as 
\begin{align}
\mathcal{I}\! =\! \! \! \int_0^\infty \! \! \! \! \sum_{m,n=0}^\infty\! \! \! \! \exp\! \! \left(\! \! -\tfrac{\rho^2}{\omega_0^2} \!\right)\! \! \tfrac{3.83 \rho D (-1)^m \left(\! \tfrac{3.83 \times 2 \rho D}{1.22 \lambda F}\! \right)^{2m} \! \! \! \left(\! \tfrac{\rho r}{\omega_0^2}\! \right)^{2n} \! \! \!}{1.22 \lambda F 4^m m! (m+1)! (n!)^2} d\rho ,\!\!
\end{align}
or equivalently
\vspace{-0.4cm}
\begin{align}
\mathcal{I}\! =\! \sum_{m,n=0}^\infty\! \tfrac{3.83 D (-1)^m \left(\! \tfrac{3.83 \times 2 D}{1.22 \lambda F}\! \right)^{2m} \left(\! \tfrac{2r}{\omega_0^2}\! \right)^{2n}}{1.22 \lambda F 4^{m+n} m! (m+1)! (n!)^2} \! \mathcal{I}_1\! ,
\label{Ib}
\end{align}
with
\vspace{-0.4cm}
\begin{align}
\mathcal{I}_1\! = \int_0^\infty \! \rho^{2m+2n+1}\! \exp\! \left(\! -\tfrac{\rho^2}{\omega_0^2}\right)\! d\rho .
\label{I_1a}
\end{align}
Furthermore, by employing~\cite[eq. 2.33/10]{gradshteyn2014table}, \eqref{I_1a} can be expressed as
\vspace{-0.4cm}
\begin{align}
\mathcal{I}_1\! = \tfrac{\Gamma \left( m+n+1\right)}{2 \left(\omega_0^{-2}\right)^{m+n+1}} .
\label{I_1b}
\end{align}
Next, by substituting~\eqref{I_1b} into~\eqref{Ib}, the former can be equivalently written as
\vspace{-0.2cm}
\begin{align}
\mathcal{I}\! =\! \tfrac{3.83 D \omega_0^{2}}{1.22 \times 2 \lambda F}\! \! \! \! \sum_{m,n=0}^\infty\! \! \! \! \tfrac{(-1)^m \left(\! \tfrac{3.83 \times 2 D}{1.22 \lambda F}\! \right)^{2m} \! \! \! \left(\! \tfrac{2r}{\omega_0^2}\! \right)^{2n} \! \! \Gamma\left( m \!+ \!n \!+ \!1 \right)}{\left(4 \omega_0^{-2}\right)^{m+n} m! (m+1)! (n!)^2} .
\label{Ic}
\end{align}
Moreover, by taking into account that $\Gamma(m+n+1)=\Gamma(1) (1)_{m+n}$, $(m+1)!=\Gamma(m+2)=(2)_{m}$, $\Gamma(1)=1$, and $\left(n!\right)^2=\Gamma(1) (1)_{n}$, \eqref{Ic} can be expressed as
\begin{align}
\mathcal{I}\! =\! \tfrac{3.83 D \omega_0^{2}}{1.22 \times 2 \lambda F}\! \sum_{m,n=0}^\infty\! \tfrac{(1)_{m+n} \left(\! -\tfrac{3.83^2 D^2 \omega_0^{2}}{1.22^2 \lambda^2 F^2}\! \right)^m\! \left(\! \tfrac{r^2}{\omega_0^{2}}\! \right)^n\!}{(1)_n (2)_m m! n!} .
\label{Id}
\end{align}
Notice that~\eqref{Id} is the fifth Humbert hypergeometric series~\cite[eq. 7.2.4/10]{prudnikov1988integrals}. Hence, \eqref{Id} can be expressed as follows
\begin{align}
\mathcal{I}\! =\! \tfrac{3.83 D \omega_0^{2}}{1.22 \times 2 \lambda F}\! \Psi_2 \left(\! 1;2,1;-\tfrac{3.83^2 D^2 \omega_0^{2}}{1.22^2 \lambda^2 F^2}\!, \tfrac{r^2}{\omega_0^{2}} \right)\! .
\label{eta_4}
\end{align}
Finally, by substituting \eqref{eta_4} into \eqref{eta_3}, the coupling efficiency can be equivalently rewritten as in \eqref{eta_2}. This concludes the~proof.

\vspace{-0.5cm}
\section*{Appendix B}
The expected value of the photon flux with respect to the misalignment fading can be evaluated as
\begin{align}
\overline{\Phi} = \int_0^\infty \Phi(r) f_r(r) dr ,
\end{align}
or equivalently,
\vspace{-0.2cm}
\begin{align}
\overline{\Phi} = \int_0^\infty k \eta(r) G_c h_l h_p(r) \tfrac{\lambda}{h c} x \tfrac{r}{\sigma_s^2}\exp\left(-\tfrac{r^2}{2\sigma_s^2}\right) dr ,
\label{av_phi}
\end{align}
where, only $\eta$ and $h_p$ are affected by the pointing error, $r$. By substituting \eqref{eta_2} and \eqref{Eq:hp1} into \eqref{av_phi}, the average emitted photon flux can be rewritten as
\begin{flalign}
\begin{aligned}
\overline{\Phi} =& k G_c h_l \tfrac{\lambda}{h c} x A_0 \tfrac{1}{\sigma_s^2} \left(\! \tfrac{3.83 \sqrt{2} D \omega_0}{1.22 \lambda F\!} \right)^2\! \\
&\times \int_0^\infty \! \! \! \! \! \! r \exp\! \left(\!-\tfrac{2 r^2}{\omega_0^2}\!-\tfrac{2 r^2}{w_{eq}^2}-\! \tfrac{r^2}{2\sigma_s^2}\!\right)\! K(r) dr ,
\end{aligned}
\end{flalign}
where
\vspace{-0.4cm}
\begin{align}
K(r) &= \left( \sum_{m=0}^\infty\! C_m \right)^2 ,
\label{sum_1}\\
C_m &= \sum_{n=0}^\infty\! \tfrac{(1)_{m+n} \left(\! -\tfrac{3.83^2 D^2 \omega_0^{2}}{1.22^2 \lambda^2 F^2}\! \right)^m\! \left(\! \tfrac{r^2}{\omega_0^{2}}\! \right)^n\!}{(1)_n (2)_m m! n!} .
\label{C_m}
\end{align}
Moreover, \eqref{sum_1} can be equivalently written as
\begin{align}
K(r) = \sum_{m,k=0}^\infty\! C_m C_k .
\label{sum_2}
\end{align}
By using \eqref{C_m}, \eqref{sum_2} can be expressed as
\begin{align}
K(r) = \sum_{m,k,n,l=0}^\infty\! \tfrac{(1)_{m+n} (1)_{k+l} \left(\! -\tfrac{3.83^2 D^2 \omega_0^{2}}{1.22^2 \lambda^2 F^2}\! \right)^{m+k}\! \left(\! \tfrac{r^2}{\omega_0^{2}}\! \right)^{n+l}\!}{(1)_n (2)_m (1)_l (2)_k m! n! k! l!} .
\label{sum_3}
\end{align}
Next, by using \eqref{sum_3}, \eqref{mean_phi_2}can be rewritten as
\begin{align}
\overline{\Phi} = k G_c h_l \tfrac{\lambda}{h c} x A_0 \tfrac{1}{\sigma_s^2} \left(\! \tfrac{3.83 \sqrt{2} D \omega_0}{1.22 \lambda F\!} \right)^2\! \Lambda(r) ,
\label{sum_4}
\end{align}
where 
\vspace{-0.4cm}
\begin{flalign}
\begin{aligned}
\Lambda(r) =\! \! \! \! \! \! \sum_{m,k,n,l=0}^\infty\! \tfrac{(1)_{m+n} (1)_{k+l} \left(\! -\tfrac{3.83^2 D^2 \omega_0^{2}}{1.22^2 \lambda^2 F^2}\! \right)^{m+k}\! \left(\! \tfrac{1}{\omega_0^{2}}\! \right)^{n+l}\!}{(1)_n (2)_m (1)_l (2)_k m! n! k! l!} \\
\times \int_0^\infty \! \! \! \! \! \! \exp\! \left(\!-\tfrac{2 r^2}{\omega_0^2}\!-\tfrac{2 r^2}{w_{eq}^2}-\! \tfrac{r^2}{2\sigma_s^2}\!\right)\! r^{2(n+l)+1} dr \! .
\label{sum_5}
\end{aligned}
\end{flalign}
After performing the integration \eqref{sum_5}, can be equivalently written as in \eqref{sum_6}, given at the top of the next page. Notice that, according to~\cite[eq. 16]{saigo1988}, \eqref{sum_6} can be expressed in terms of the general quadruple hypergeometric function as in \eqref{sum_7}, given at the top of the next page. Finally, by substituting \eqref{sum_7} into \eqref{sum_4}, the average emitted photon flux can be written as in \eqref{mean_phi_2}. This concludes the proof.

\begin{figure*}
\begin{align}
\Lambda(r) = \tfrac{1}{2} \left( \tfrac{2}{\omega_0^2}\! +\tfrac{2}{w_{eq}^2}+\! \tfrac{1}{2\sigma_s^2} \right)^{-1} \sum_{m,k,n,l=0}^\infty\! \tfrac{(1)_{m+n} (1)_{k+l} (1)_{n+l} \left(-\tfrac{3.83^2 D^2 \omega_0^{2}}{1.22^2 \lambda^2 F^2}\right)^{m+k}\! \left(\tfrac{\tfrac{1}{\omega_0^{2}}}{\tfrac{2}{\omega_0^2}\! +\tfrac{2}{w_{eq}^2}+\! \tfrac{1}{2\sigma_s^2}}\right)^{n+l}\!}{(1)_n (2)_m (1)_l (2)_k m! n! k! l!} .
\label{sum_6}
\end{align}
\line(1,0){512}
\end{figure*}

\begin{figure*}
\begin{flalign}
\begin{aligned}
\Lambda(r) = \tfrac{1}{2} \left( \tfrac{2}{\omega_0^2}\! +\tfrac{2}{w_{eq}^2}+\! \tfrac{1}{2\sigma_s^2} \right)^{-1} F^{(4)} \left[
\begin{array}{c}
\!\!\text{-}\!\!::: \!\!\text{-}; \!\!\text{-}; \!\!\text{-}; \!\!\text{-}: \!\!\text{-}; 1; \!\!\text{-}; 1; \!\!\text{-}; 1: \!\!\text{-}; \!\!\text{-}; \!\!\text{-}; \!\!\text{-}; \\
\!\!\text{-}\!\!::: \!\!\text{-}; \!\!\text{-}; \!\!\text{-}; \!\!\text{-}: \!\!\text{-}; \!\!\text{-}; \!\!\text{-}; \!\!\text{-}; \!\!\text{-}; \!\!\text{-}: 2;  2;  1;  1;
\end{array} 
\!\!\! \text{-}\tfrac{3.83^2 D^2 \omega_0^{2}}{1.22^2 \lambda^2 F^2},\!\!\text{-}\tfrac{3.83^2 D^2 \omega_0^{2}}{1.22^2 \lambda^2 F^2},\!\!\tfrac{\tfrac{1}{\omega_0^{2}}}{\tfrac{2}{\!\omega_0^2}\! +\!\tfrac{2}{w_{eq}^2}\!+\! \tfrac{1}{2\sigma_s^2}\!},\!\!\tfrac{\tfrac{1}{\omega_0^{2}}}{\tfrac{2}{\!\omega_0^2}\! +\!\tfrac{2}{w_{eq}^2}\!+\! \tfrac{1}{2\sigma_s^2}\!}\! \right] .
\label{sum_7}
\end{aligned}
\end{flalign}
\line(1,0){512}
\end{figure*}

\vspace{-0.3cm}
\bibliographystyle{IEEEtran}
\bibliography{bib}

\begin{thebibliography}{10}
\providecommand{\url}[1]{#1}
\csname url@samestyle\endcsname
\providecommand{\newblock}{\relax}
\providecommand{\bibinfo}[2]{#2}
\providecommand{\BIBentrySTDinterwordspacing}{\spaceskip=0pt\relax}
\providecommand{\BIBentryALTinterwordstretchfactor}{4}
\providecommand{\BIBentryALTinterwordspacing}{\spaceskip=\fontdimen2\font plus
\BIBentryALTinterwordstretchfactor\fontdimen3\font minus
  \fontdimen4\font\relax}
\providecommand{\BIBforeignlanguage}[2]{{%
\expandafter\ifx\csname l@#1\endcsname\relax
\typeout{** WARNING: IEEEtran.bst: No hyphenation pattern has been}%
\typeout{** loaded for the language `#1'. Using the pattern for}%
\typeout{** the default language instead.}%
\else
\language=\csname l@#1\endcsname
\fi
#2}}
\providecommand{\BIBdecl}{\relax}
\BIBdecl

\bibitem{dombrowski2018toward}
T.~Dombrowski, V.~Rankovic, and T.~Moser, ``Toward the optical cochlear
  implant,'' \emph{Cold Spring Harb. Perspect. Med.}, p. a033225, Oct. 2018.

\bibitem{wilson2008implants}
B.~S. Wilson and M.~F. Dorman, ``Cochlear implants: {A} remarkable past and a
  brilliant future,'' \emph{Hear. Res.}, vol. 242, no. 1-2, pp. 3--21, Aug.
  2008.

\bibitem{zeng2008implants}
F.~G. Zeng, S.~Rebscher, W.~Harrison, X.~Sun, and H.~Feng, ``Cochlear implants:
  {System} design, integration, and evaluation,'' \emph{IEEE Rev. Biomed.
  Eng.}, vol.~1, pp. 115--142, Nov. 2008.

\bibitem{wilson2017modern}
B.~S. Wilson, ``The modern cochlear implant: {A} triumph of biomedical
  engineering and the first substantial restoration of human sense using a
  medical intervention,'' \emph{IEEE Pulse}, vol.~8, no.~2, pp. 29--32, Mar.
  2017.

\bibitem{Moser2020}
T.~Moser and A.~Dieter, ``Towards optogenetic approaches for hearing
  restoration,'' \emph{Biochemical and Biophysical Research Communications},
  feb 2020.

\bibitem{Hernandez2014}
V.~H. Hernandez, A.~Gehrt, K.~Reuter, Z.~Jing, M.~Jeschke, A.~M. Schulz,
  G.~Hoch, M.~Bartels, G.~Vogt, C.~W. Garnham, H.~Yawo, Y.~Fukazawa, G.~J.
  Augustine, E.~Bamberg, S.~Kügler, T.~Salditt, L.~de~Hoz, N.~Strenzke, and
  T.~Moser, ``Optogenetic stimulation of the auditory pathway,'' \emph{J. Clin.
  Invest.}, vol. 124, no.~3, pp. 1114--1129, Feb. 2014.

\bibitem{wrobel2018optogenetic}
C.~Wrobel, A.~Dieter, A.~Huet, D.~Keppeler, C.~J. Duque-Afonso, C.~Vogl,
  G.~Hoch, M.~Jeschke, and T.~Moser, ``Optogenetic stimulation of cochlear
  neurons activates the auditory pathway and restores auditory-driven behavior
  in deaf adult gerbils,'' \emph{Sci. Transl. Med.}, vol.~10, no. 449, p.
  eaao0540, Jul. 2018.

\bibitem{mager2018high}
T.~Mager, D.~L. de~la Morena, V.~Senn, J.~Schlotte, K.~Feldbauer, C.~Wrobel,
  S.~Jung, K.~Bodensiek, V.~Rankovic, L.~Browne \emph{et~al.}, ``High frequency
  neural spiking and auditory signaling by ultrafast red-shifted
  optogenetics,'' \emph{Nat. Commun.}, vol.~9, no.~1, p. 1750, May 2018.

\bibitem{keppeler2018ultrafast}
D.~Keppeler, R.~M. Merino, D.~L. de~la Morena, B.~Bali, A.~T. Huet, A.~Gehrt,
  C.~Wrobel, S.~Subramanian, T.~Dombrowski, F.~Wolf \emph{et~al.}, ``Ultrafast
  optogenetic stimulation of the auditory pathway by targeting-optimized
  chronos,'' \emph{The EMBO journal}, vol.~37, no.~24, p. e99649, Nov. 2018.

\bibitem{Xu2018}
Y.~Xu, N.~Xia, M.~Lim, X.~Tan, M.~H. Tran, E.~Boulger, F.~Peng, H.~Young,
  C.~Rau, A.~Rack, and C.-P. Richter, ``Multichannel optrodes for photonic
  stimulation,'' \emph{Neurophotonics}, vol.~5, no.~04, p.~1, Oct. 2018.

\bibitem{Hart2019}
W.~L. Hart, T.~Kameneva, A.~K. Wise, and P.~R. Stoddart, ``Biological
  considerations of optical interfaces for neuromodulation,'' \emph{Adv. Opt.
  Mater.}, vol.~7, no.~19, p. 1900385, Jul. 2019.

\bibitem{trevlakis2019optical}
S.~E. Trevlakis, A.-A.~A. Boulogeorgos, P.~C. Sofotasios, S.~Muhaidat, and
  G.~K. Karagiannidis, ``Optical wireless cochlear implants,'' \emph{Biomed.
  Opt. Express}, vol.~10, no.~2, pp. 707--730, Nov. 2019.

\bibitem{ackermann2008designing}
D.~M. Ackermann~Jr, B.~Smith, X.-F. Wang, K.~L. Kilgore, and P.~H. Peckham,
  ``Designing the optical interface of a transcutaneous optical telemetry
  link,'' \emph{IEEE Trans. Biomed. Eng.}, vol.~55, no.~4, pp. 1365--1373, Apr.
  2008.

\bibitem{gil2012feasibility}
Y.~Gil, N.~Rotter, and S.~Arnon, ``Feasibility of retroreflective transdermal
  optical wireless communication,'' \emph{Appl. Opt.}, vol.~51, no.~18, pp.
  4232--4239, Jun. 2012.

\bibitem{liu2015bidirectional}
T.~Liu, J.~Anders, and M.~Ortmanns, ``Bidirectional optical transcutaneous
  telemetric link for brain machine interface,'' \emph{Electron. Lett.},
  vol.~51, no.~24, pp. 1969--1971, Nov. 2015.

\bibitem{wells2014laser}
J.~D. Wells, A.~Xing, M.~P. Bendett, M.~D. Keller, B.~J. Norton, J.~M. Owen,
  S.~Yuan, R.~W. Royse, and C.~A. Lemaire, ``Laser-based nerve stimulators for,
  eg, hearing restoration in cochlear prostheses and method,'' Jul.~29 2014, uS
  Patent 8.792.978.

\bibitem{nguyen2017fabrication}
H.~Nguyen, M.~M.~P. Arnob, A.~T. Becker, J.~C. Wolfe, M.~K. Hogan, P.~J.
  Horner, and W.-C. Shih, ``Fabrication of multipoint side-firing optical fiber
  by laser micro-ablation,'' \emph{Opt. Lett.}, vol.~42, no.~9, pp. 1808--1811,
  Apr. 2017.

\bibitem{pisanello2014multipoint}
F.~Pisanello, L.~Sileo, I.~A. Oldenburg, M.~Pisanello, L.~Martiradonna, J.~A.
  Assad, B.~L. Sabatini, and M.~D. Vittorio, ``Multipoint-emitting optical
  fibers for spatially addressable in~vivo optogenetics,'' \emph{Neuron},
  vol.~82, no.~6, pp. 1245--1254, Jun. 2014.

\bibitem{pisanello2018tailoring}
M.~Pisanello, F.~Pisano, L.~Sileo, E.~Maglie, E.~Bellistri, B.~Spagnolo,
  G.~Mandelbaum, B.~L. Sabatini, M.~De~Vittorio, and F.~Pisanello, ``Tailoring
  light delivery for optogenetics by modal demultiplexing in tapered optical
  fibers,'' \emph{Sci. Rep.}, vol.~8, no.~1, p. 4467, Oct. 2018.

\bibitem{pisano2018focused}
F.~Pisano, M.~Pisanello, L.~Sileo, A.~Qualtieri, B.~L. Sabatini,
  M.~De~Vittorio, and F.~Pisanello, ``Focused ion beam nanomachining of tapered
  optical fibers for patterned light delivery,'' \emph{Microelectron. Eng.},
  vol. 195, pp. 41--49, Mar. 2018.

\bibitem{prudnikov1988integrals}
A.~P. Prudnikov, Y.~A. Bry\v{c}kov, and O.~I. Mari\v{c}ev, \emph{Integrals and
  Series of Special Functions}.\hskip 1em plus 0.5em minus 0.4em\relax Moscow,
  Russia: Science, 1983.

\bibitem{saigo1988}
M.~Saigo, ``On properties of hypergeometric functions of three variables,
  {F${}_M$} and {F${}_G$},'' \emph{Rendiconti del Circolo Matematico di
  Palermo}, vol.~37, no.~3, pp. 449--468, Sep. 1988.

\bibitem{Oltulu2018}
P.~Oltulu, B.~Ince, N.~Kokbudak, S.~Findik, and F.~Kilinc, ``Measurement of
  epidermis, dermis, and total skin thicknesses from six different body regions
  with a new ethical histometric technique,'' \emph{Turkish Journal of Plastic
  Surgery}, vol.~26, no.~2, p.~56, 2018.

\bibitem{Zou2015}
Y.~Zou, W.~Zhang, F.~Tian, F.~S. Chau, and G.~Zhou, ``Development of miniature
  tunable multi-element alvarez lenses,'' \emph{{IEEE} J. Sel. Top. Quantum
  Electron.}, vol.~21, no.~4, pp. 100--107, Jul. 2015.

\bibitem{Hwang2017}
K.~Hwang, Y.-H. Seo, and K.-H. Jeong, ``Microscanners for optical
  endomicroscopic applications,'' \emph{Micro and Nano Syst. Lett.}, vol.~5,
  no.~1, Jan. 2017.

\bibitem{Zhou2017}
G.~Zhou and C.~Lee, Eds., \emph{Optical {MEMS}, Nanophotonics, and Their
  Applications}.\hskip 1em plus 0.5em minus 0.4em\relax {CRC} Press, Dec. 2017.

\bibitem{Khoshnoud2012}
F.~Khoshnoud and C.~W. de~Silva, ``Recent advances in {MEMS} sensor technology
  {\textendash} biomedical applications,'' \emph{{IEEE} Instrumentation {\&}
  Measurement Magazine}, vol.~15, no.~1, pp. 8--14, feb 2012.

\bibitem{Sezen2005}
A.~S. Sezen, S.~Sivaramakrishnan, S.~Hur, R.~Rajamani, W.~Robbins, and B.~J.
  Nelson, ``Passive wireless {MEMS} microphones for biomedical applications,''
  \emph{Journal of Biomechanical Engineering}, vol. 127, no.~6, pp. 1030--1034,
  jul 2005.

\bibitem{Marcuse1978}
D.~Marcuse, ``Gaussian approximation of the fundamental modes of graded-index
  fibers,'' \emph{J. Opt. Soc. Am.}, vol.~68, no.~1, p. 103, Jan. 1978.

\bibitem{Srinivasan2010}
A.~G. Srinivasan, D.~M. Landsberger, and R.~V. Shannon, ``Current focusing
  sharpens local peaks of excitation in cochlear implant stimulation,''
  \emph{Hearing Research}, vol. 270, no. 1-2, pp. 89--100, dec 2010.

\bibitem{Friesen2001}
L.~M. Friesen, R.~V. Shannon, D.~Baskent, and X.~Wang, ``Speech recognition in
  noise as a function of the number of spectral channels: Comparison of
  acoustic hearing and cochlear implants,'' \emph{The Journal of the Acoustical
  Society of America}, vol. 110, no.~2, pp. 1150--1163, aug 2001.

\bibitem{Kong2004}
Y.-Y. Kong, R.~Cruz, J.~A. Jones, and F.-G. Zeng, ``Music perception with
  temporal cues in acoustic and electric hearing,'' \emph{Ear and Hearing},
  vol.~25, no.~2, pp. 173--185, apr 2004.

\bibitem{zhou2006optic}
K.~Zhou, L.~Zhang, X.~Chen, and I.~Bennion, ``Optic sensors of high
  refractive-index responsivity and low thermal cross sensitivity that use
  fiber bragg gratings of $> 80^{\circ}$ tilted structures,'' \emph{Opt.
  Lett.}, vol.~31, no.~9, pp. 1193--1195, May 2006.

\bibitem{cotillard2014regeneration}
R.~Cotillard, G.~Laffont, and P.~Ferdinand, ``Regeneration of tilted fiber
  bragg gratings,'' in \emph{23rd International Conference on Optical Fibre
  Sensors}, Jun. 2014, p. 91572S.

\bibitem{bharathan2018fiber}
G.~Bharathan, D.~D. Hudson, R.~I. Woodward, S.~D. Jackson, and A.~Fuerbach,
  ``In-fiber polarizer based on a 45-degree tilted fluoride fiber bragg grating
  for mid-infrared fiber laser technology,'' \emph{OSA Continuum}, vol.~1,
  no.~1, pp. 56--63, Sep. 2018.

\bibitem{Mou2009}
C.~Mou, K.~Zhou, L.~Zhang, and I.~Bennion, ``Characterization of
  45${}^{\circ}$-tilted fiber grating and its polarization function in fiber
  ring laser,'' \emph{J. Opt. Soc. Am. B}, vol.~26, no.~10, p. 1905, Sep. 2009.

\bibitem{trevlakis2018signal}
S.~Trevlakis, A.-A. Boulogeorgos, and G.~Karagiannidis, ``Signal quality
  assessment for transdermal optical wireless communications under pointing
  errors,'' \emph{Technologies}, vol.~6, no.~4, p. 109, 2018.

\bibitem{Lister2012}
T.~Lister, P.~A. Wright, and P.~H. Chappell, ``Optical properties of human
  skin,'' \emph{Journal of Biomedical Optics}, vol.~17, no.~9, p. 0909011, sep
  2012.

\bibitem{Bosschaart2011}
N.~Bosschaart, R.~Mentink, J.~H. Kok, T.~G. van Leeuwen, and M.~C.~G. Aalders,
  ``Optical properties of neonatal skin measured in vivo as a function of age
  and skin pigmentation,'' \emph{Journal of Biomedical Optics}, vol.~16, no.~9,
  p. 097003, 2011.

\bibitem{bashkatov2011optical}
A.~N. Bashkatov, E.~A. Genina, and V.~V. Tuchin, ``Optical properties of skin,
  subcutaneous, and muscle tissues: a review,'' \emph{J. Innov. Opt. Health
  Sci.}, vol.~4, no.~01, pp. 9--38, Jan. 2011.

\bibitem{bashkatov2005optical}
A.~Bashkatov, E.~Genina, V.~Kochubey, and V.~Tuchin, ``Optical properties of
  human skin, subcutaneous and mucous tissues in the wavelength range from 400
  to 2000 nm,'' \emph{J. Phys. D: Appl. Phys.}, vol.~38, no.~15, p. 2543, Jul.
  2005.

\bibitem{chan1996effects}
E.~K. Chan, B.~Sorg, D.~Protsenko, M.~O'Neil, M.~Motamedi, and A.~J. Welch,
  ``Effects of compression on soft tissue optical properties,'' \emph{IEEE J.
  Sel. Topics Quantum Electron.}, vol.~2, no.~4, pp. 943--950, Dec. 1996.

\bibitem{van1989skin}
M.~Van~Gemert, S.~L. Jacques, H.~Sterenborg, and W.~Star, ``Skin optics,''
  \emph{IEEE Trans. Biomed. Eng.}, vol.~36, no.~12, pp. 1146--1154, Dec. 1989.

\bibitem{A:Outage_Capacity_for_FSO_with_pointing_errors}
A.~A. Farid and S.~Hranilovic, ``Outage capacity optimization for free-space
  optical links with pointing errors,'' \emph{J. Lightwave Technol.}, vol.~25,
  no.~7, pp. 1702--1710, Jul. 2007.

\bibitem{A:BER_performance_of_FSO_link_over_strong_atm_turbulence_channels_with_pointing_errors}
H.~G. Sandalidis, T.~A. Tsiftsis, G.~K. Karagiannidis, and M.~Uysal, ``{BER}
  performance of {FSO} links over strong atmospheric turbulence channels with
  pointing errors,'' \emph{IEEE Commun. Lett.}, vol.~12, no.~1, pp. 44--46,
  Jan. 2008.

\bibitem{arnon2003Effects}
S.~Arnon, ``Effects of atmospheric turbulence and building sway on optical
  wireless-communication systems,'' \emph{Opt. Lett.}, vol.~28, no.~2, pp.
  129--131, Jan. 2003.

\bibitem{B:Probability_Random_Variables_and_Stochastic_Processes}
A.~Papoulis and S.~Pillai, \emph{Probability, Random Variables, and Stochastic
  Processes}, ser. McGraw-Hill series in electrical engineering: Communications
  and signal processing.\hskip 1em plus 0.5em minus 0.4em\relax Tata
  McGraw-Hill, Jan. 2002.

\bibitem{sabry2017plane}
Y.~Sabry, D.~Khalil, B.~Saadany, and T.~Bourouina, ``In-plane optical beam
  collimation using a three-dimensional curved {MEMS} mirror,''
  \emph{Micromachines}, vol.~8, no.~5, p. 134, Apr. 2017.

\bibitem{yu2009modeling}
H.~Yu, S.~Wang, J.~Fu, M.~Qiu, Y.~Li, F.~Gu, and L.~Tong, ``Modeling bending
  losses of optical nanofibers or nanowires,'' \emph{Appl. Opt.}, vol.~48,
  no.~22, pp. 4365--4369, Jul. 2009.

\bibitem{Wang2016}
G.~Wang, C.~Wang, Z.~Yan, and L.~Zhang, ``Highly efficient spectrally encoded
  imaging using a 45{\textdegree} tilted fiber grating,'' \emph{Opt. Lett.},
  vol.~41, no.~11, p. 2398, May 2016.

\bibitem{Hadfield2009}
R.~H. Hadfield, ``Single-photon detectors for optical quantum information
  applications,'' \emph{Nat. Photonics}, vol.~3, no.~12, pp. 696--705, Dec.
  2009.

\bibitem{Sjulson2007}
L.~Sjulson and G.~Miesenböck, ``Optical recording of action potentials and
  other discrete physiological events: A perspective from signal detection
  theory,'' \emph{Physiology}, vol.~22, no.~1, pp. 47--55, feb 2007.

\bibitem{Hendel2008}
T.~Hendel, M.~Mank, B.~Schnell, O.~Griesbeck, A.~Borst, and D.~F. Reiff,
  ``Fluorescence changes of genetic calcium indicators and {OGB}-1 correlated
  with neural activity and calcium in vivo and in vitro,'' \emph{Journal of
  Neuroscience}, vol.~28, no.~29, pp. 7399--7411, jul 2008.

\bibitem{Sasaki2008}
T.~Sasaki, N.~Takahashi, N.~Matsuki, and Y.~Ikegaya, ``Fast and accurate
  detection of action potentials from somatic calcium fluctuations,''
  \emph{Journal of Neurophysiology}, vol. 100, no.~3, pp. 1668--1676, sep 2008.

\bibitem{Vogelstein2009}
J.~T. Vogelstein, B.~O. Watson, A.~M. Packer, R.~Yuste, B.~Jedynak, and
  L.~Paninski, ``Spike inference from calcium imaging using sequential monte
  carlo methods,'' \emph{Biophysical Journal}, vol.~97, no.~2, pp. 636--655,
  jul 2009.

\bibitem{wilt2013photon}
B.~A. Wilt, J.~E. Fitzgerald, and M.~J. Schnitzer, ``Photon shot noise limits
  on optical detection of neuronal spikes and estimation of spike timing,''
  \emph{Biophys. J.}, vol. 104, no.~1, pp. 51--62, Jan. 2013.

\bibitem{Han2012}
X.~Han, ``In vivo application of optogenetics for neural circuit analysis,''
  \emph{{ACS} Chem. Neurosci.}, vol.~3, no.~8, pp. 577--584, Jul. 2012.

\bibitem{ICNIRP2013a}
{ICNIRP}, ``{ICNIRP} guidelines on limits of exposure to laser radiation of
  wavelengths between 180 nm and 1,000 $\mu$m,'' \emph{Health Physics}, vol.
  105, no.~3, pp. 271--295, Sep. 2013.

\bibitem{Jahns2008}
J.~Jahns, Q.~Cao, and S.~Sinzinger, ``Micro- and nanooptics - an overview,''
  \emph{Laser {\&} Photonics Review}, vol.~2, no.~4, pp. 249--263, aug 2008.

\bibitem{Yang2004}
H.-M. Yang, S.-Y. Huang, C.-W. Lee, T.-S. Lay, and W.-H. Cheng, ``High-coupling
  tapered hyperbolic fiber microlens and taper asymmetry effect,''
  \emph{Journal of Lightwave Technology}, vol.~22, no.~5, pp. 1395--1401, may
  2004.

\bibitem{Toyoshima2006}
M.~Toyoshima, ``Maximum fiber coupling efficiency and optimum beam size in the
  presence of random angular jitter for free-space laser systems and their
  applications,'' \emph{J. Opt. Soc. Am. A}, vol.~23, no.~9, p. 2246, Sep.
  2006.

\bibitem{gradshteyn2014table}
I.~S. Gradshteyn and I.~M. Ryzhik, \emph{Table of integrals, series, and
  products}, 7th~ed.\hskip 1em plus 0.5em minus 0.4em\relax Elsevier/Academic
  Press, Amsterdam, 2007.

\end{thebibliography}

\vspace{-1cm}
\begin{IEEEbiography}
	[{\includegraphics[width=1in,height=1.25in,clip,keepaspectratio]{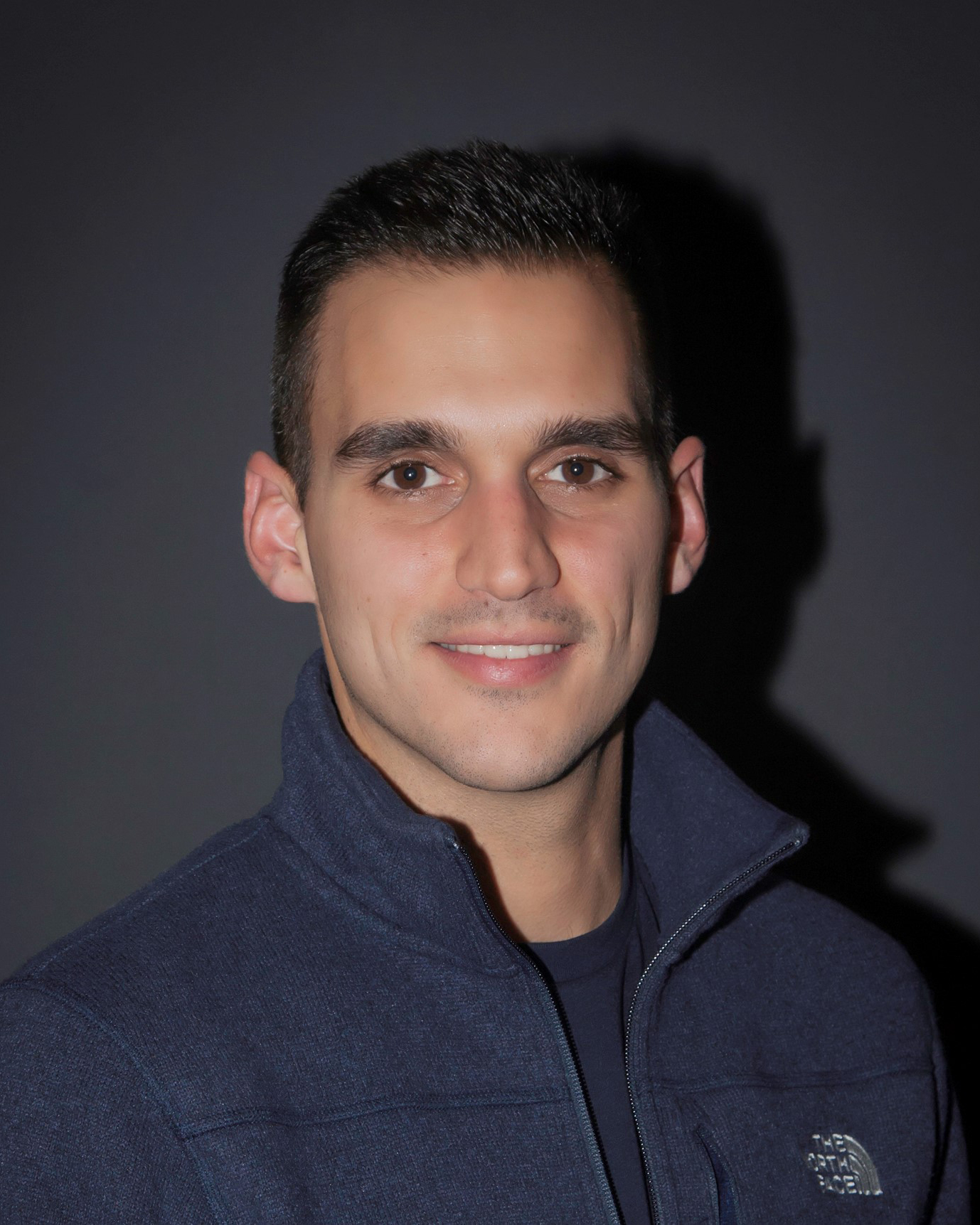}}]
	{Stylianos E. Trevlakis} was born in Thessaloniki, Greece in 1991. He received the Electrical and Computer Engineering (ECE) diploma (5 year) from the Aristotle University of Thessaloniki (AUTh) in 2016. Afterwards he served in the Hellenic Army in for nine months in the Research Office as well as at the Office of Research and Informatics of the School of Management and Officers. During 2017, he joined the Information Technologies Institute, while from October 2017, he joined WCSG underthe leadership Prof. Karagiannidis in AUTh as a PhD candidate. 
	
	His research interests are in the area of Wireless Communications, with emphasis on Optical Wireless Communications, and Communications \& Signal Processing for Biomedical Engineering.
\end{IEEEbiography}

\vspace{-1cm}
\begin{IEEEbiography}
	[{\includegraphics[width=1in,height=1.25in,clip,keepaspectratio]{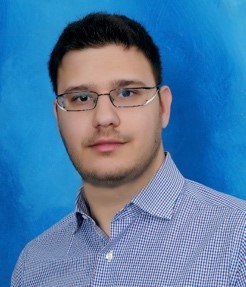}}]
	{Alexandros--Apostolos A. Boulogeorgos} (S’11-M’16-SM’19) was born in Trikala, Greece in 1988. He received the Electrical and Computer Engineering (ECE) diploma (5 year) and Ph.D. degree in Wireless Communications from the Aristotle University of Thessaloniki (AUTh) in 2012 and 2016, respectively. From November 2012, he has been a member of the wireless communications system group of AUTh. During 2017, he joined the information technologies institute, while from November 2017, he has joined the Department of Digital Systems, University of Piraeus. Moreover, from October 2012 until September 2016, he was a teaching assistant at the department of ECE of AUTh, whereas, from February 2017, he serves as an adjunct lecturer at the Department of ECE of the University of Western Macedonia and as an visiting lecturer at the Department of Computer Science and Biomedical Informatics of the University of Thessaly.
	
	He has been involved as member of Technical Program Committees in several IEEE and non-IEEE conferences and served as a reviewer in various IEEE journals and conferences. He was awarded with the “Distinction Scholarship Award” of the Research Committee of AUTh for the year 2014, was recognized as an exemplary reviewer for IEEE Communication Letters for 2016 (top $3\%$ of reviewers) and was named a top peer reviewer ($<1\%$) in Cross-Field and Computer Science in the Global Peer Review Awards 2019, which was presented by the Web of Science and Publons. 
 
	His current research interests spans in the area of wireless communications and networks with emphasis in high frequency communications, optical wireless communications and communications for biomedical applications. 
\end{IEEEbiography}

\vspace{-1cm}
\begin{IEEEbiography}
	[{\includegraphics[width=1in,height=1.25in,clip,keepaspectratio]{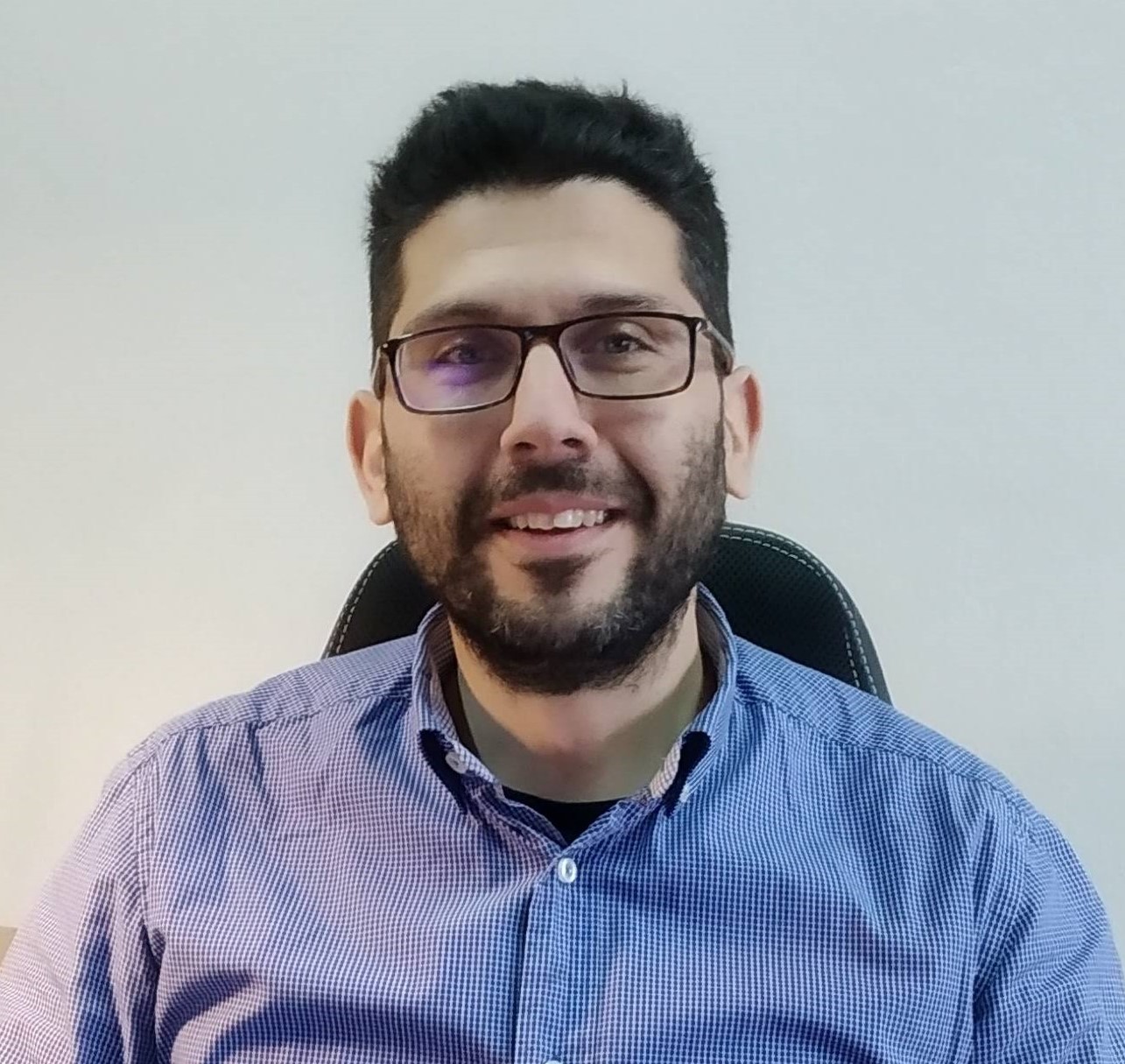}}]
	{Nestor D. Chatzidiamantis} (S’08, M’14) was born in Los Angeles, CA, USA, in 1981. He received the Diploma degree (5 years) in electrical and computer engineering (ECE) from the Aristotle University of Thessaloniki (AUTH), Greece, in 2005, the M.Sc. degree in telecommunication networks and software from the University of Surrey, U.K., in 2006, and the Ph.D. degree from the ECE Department, AUTH, in 2012. From 2012 through 2015, he worked as a Post-Doctoral Research Associate in AUTH and from 2016 to 2018, as a Senior Engineer at the Hellenic Electricity Distribution Network Operator (HEDNO). Since 2018, he has been Assistant Professor at the ECE Department of AUTH and member of the Telecommunications Laboratory.
	
	His research areas span signal processing techniques for communication systems, performance analysis of wireless communication systems over fading channels, communications theory, cognitive radio and free-space optical communications.
\end{IEEEbiography}

\vspace{-1cm}
\begin{IEEEbiography}
	[{\includegraphics[width=1in,height=1.25in,clip,keepaspectratio]{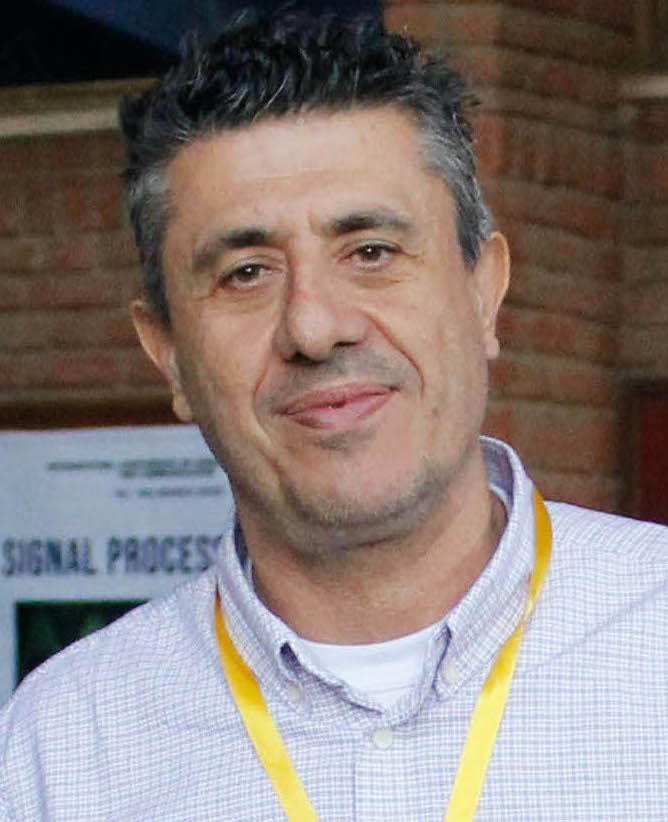}}]
	{George K. Karagiannidis} (M’96-SM’03-F’14) was born in Pithagorion, Samos Island, Greece. He received the University Diploma (5 years) and PhD degree, both in electrical and computer engineering from the University of Patras, in 1987 and 1999, respectively. From 2000 to 2004, he was a Senior Researcher at the Institute for Space Applications and Remote Sensing, National Observatory of Athens, Greece. In June 2004, he joined the faculty of Aristotle University of Thessaloniki, Greece where he is currently Professor in the Electrical \& Computer Engineering Dept. and Head of Wireless Communications Systems Group (WCSG).  He is also Honorary Professor at South West Jiaotong University, Chengdu, China.

	His research interests are in the broad area of Digital Communications Systems and Signal processing, with emphasis on Wireless Communications, Optical Wireless Communications, Wireless Power Transfer and Applications and Communications \& Signal Processing for Biomedical Engineering.

	Dr. Karagiannidis has been involved as General Chair, Technical Program Chair and member of Technical Program Committees in several IEEE and non-IEEE conferences. In the past, he was Editor in several IEEE journals and from 2012 to 2015 he was the Editor-in Chief of IEEE Communications Letters. Currently, he serves as Associate Editor-in Chief of IEEE Open Journal of Communications Society.

	Dr. Karagiannidis is one of the highly-cited authors across all areas of Electrical Engineering, recognized from Clarivate Analytics as Web-of-Science Highly-Cited Researcher in the five consecutive years 2015-2019.
\end{IEEEbiography}

\end{document}